\documentclass[journal,12pt,onecolumn,draftclsnofoot,]{IEEEtran}
\usepackage{adjustbox}
\usepackage[utf8]{inputenc}
\usepackage[T1]{fontenc}
\usepackage{graphicx}

\usepackage{comment}
\usepackage{float}
\usepackage{csquotes}
\usepackage{enumitem}
\usepackage[normalem]{ulem}
\usepackage{csquotes}

\usepackage{array, makecell, multirow, booktabs}
\setcellgapes{2pt}
\usepackage{amsmath, amsthm, amsfonts}
\usepackage{amssymb}
\usepackage{color, xcolor}

\usepackage{tikz}
\usepackage{comment}
 \usepackage{url}
\usepackage{array}
\usepackage{longtable}
\usepackage{hyperref}
\usepackage{booktabs}
\usepackage{booktabs,multirow,xcolor}
\usepackage{graphicx}   
\renewcommand{\arraystretch}{1.3} 

\usepackage[linesnumbered, algoruled, boxed, lined]{algorithm2e}
\usepackage{graphicx}

\usepackage{color,soul}

\usepackage[table,xcdraw]{xcolor}
\usepackage{array}
\usepackage{graphicx}
\usepackage{subcaption}

\usepackage{booktabs, multirow, multicol, tabularx}
\usepackage[table]{xcolor}
\definecolor{yesgreen}{RGB}{198,239,206}
\definecolor{nored}{RGB}{255,199,206}
\definecolor{nagrey}{RGB}{224,224,224}
\definecolor{overlow}{RGB}{184,238,184}
\definecolor{overhigh}{RGB}{255,179,179}
\definecolor{overvm}{RGB}{255,224,178}

\begin{document}

\bstctlcite{IEEEexample:BSTcontrol}

\title{K8S Power Irrigation: Deep Reinforcement Learning for Performance-Aware Power Efficiency of Kubernetes Cloud-Native Microservices}

   \author{
	\IEEEauthorblockN{
		Zouhir Bellal,
		Laaziz Lahlou, 
		Nadjia Kara,
        Timothy Murphy, Tan Phat Nguyen
	}
\thanks{This work has been submitted to the IEEE for possible publication. Copyright may be transferred without notice, after which this version may no longer be accessible.}
    \thanks{This work is supported by  NSERC and Ericsson Canada, Grant \# 561771}
    \thanks{Z. Bellal is with  École de technologie supérieure (ÉTS), Montréal, Canada (zouhir.bellal.1@ens.etsmtl.ca)}
    \thanks{L. Lahlou is with  École de technologie supérieure (ÉTS), Montréal, Canada (laaziz.lahlou@etsmtl.ca)}
    \thanks{N. Kara is with  École de technologie supérieure (ÉTS), Montréal, Canada (nadjia.kara@etsmtl.ca)}
    \thanks{T. Murphy is with Ericsson, Canada (timothy.murphy@ericsson.com)}
    \thanks{T. Phat Nguyen is with Ericsson, Canada (Tan.Phat.Nguyen@ericsson.com)}
   
} 

\markboth{This work has been submitted to the IEEE for possible publication}{}

\maketitle

\begin{abstract}

Modern cloud platforms are facing a sharp increase in power demand driven by the rapid adoption of AI-powered applications, making power optimization urgent under net-zero commitments and sustainability goals. Yet, reducing power in production remains challenging for latency-sensitive microservices (e.g., backend pipelines for autonomous driving, real-time AI inference, and AI/IoT workloads), where performance violations directly impact user experience and operational risk. These microservices exhibit heterogeneous workload characteristics (CPU-bound, memory-bound, and mixed) and diverse load patterns. In multi-tenant environments, especially for memory-intensive workloads, contention on shared uncore resources (e.g., last-level cache and memory bandwidth) can degrade performance and trigger violations of performance requirements. As a conservative safeguard, providers often pin servers to performance mode (maximum core and uncore frequencies). This occurs because existing power governors largely disregard application-level performance requirements and ignore uncore interference under contention, maintaining reliability by sacrificing energy efficiency and causing systematic power over-provisioning. To address this, we introduce K8SPI (Kubernetes Power Irrigation), a hierarchical reinforcement learning (HRL) controller that jointly optimizes CPU core and uncore frequencies for cloud-native deployments. K8SPI uses a two-stage control architecture: a coarse-grained agent rapidly mitigates performance violations, while a fine-grained agent iteratively minimizes power consumption once performance requirements are met. Guided by multi-level telemetry spanning hardware, Kubernetes, and application performance signals, K8SPI adapts to workload heterogeneity and cross-microservice interference to meet performance requirements with minimal node power. We evaluate K8SPI on a Kubernetes testbed across multiple scenarios. Experimental results show that K8SPI reduces node-level power consumption by 23--30\% relative to the Linux performance governor while keeping performance requirement violations below 2--3\%, even under severe uncore contention and dynamic load fluctuations.
\end{abstract}


\IEEEpeerreviewmaketitle

\section{Introduction}
\label{introduction}
By 2030, the electricity demand of the IT sector is projected to reach approximately 3,200 TWh~\cite{electricityDemand}. These projections likely underestimate future consumption due to the rapid proliferation of energy-intensive Artificial Intelligence (AI) workloads, particularly generative AI. Reports indicate that a single ChatGPT query may consume significantly more energy than a conventional web search~\cite{brusseltimes, katesaenko}, with the additional global power capacity required for AI projected to reach 327 GW by 2030~\cite{pilz2025AIpower}. Even with innovations such as DeepSeek that reduce the cost of developing large language models (LLMs), overall energy demand is unlikely to slow down due to large-scale deployment and workload growth~\cite{deKruijff2025deepseek}.

This surge, combined with regulatory pressures such as carbon pricing schemes~\cite{ahuja2024legal}, prompts energy efficiency to be a primary operational objective for cloud providers. However, achieving efficiency in cloud-native environments, characterized by multi-tenancy, heterogeneous microservices, and dynamic load, remains a complex control problem.

Dynamic Voltage and Frequency Scaling (DVFS) is the standard mechanism for managing processor power. Modern architectures support scaling for both processing cores and the ``uncore'' (interconnects and Last Level Cache). While core frequency dictates computational throughput, uncore frequency governs memory subsystem performance, including L3 cache latency and bandwidth availability. Existing power governors (e.g., intel\_pstate, lunix schedutil) rely on hardware-derived utilization metrics (CPU usage, instruction by cycle) to adjust these frequencies. However, these governors optimize system-level utilization rather than enforcing microservice-level performance requirements (e.g., latency constraints). Consequently, they may over-provision resources—leading to unnecessary power waste—or under-provision them, resulting in Performance requirement violations. This limitation becomes particularly critical in performance-sensitive microservices deployed in cloud environments.

The problem is exacerbated in multi-tenant nodes where microservices exhibit diverse power--performance sensitivities. CPU-bound services rely on core frequency, while memory-bound services depend heavily on uncore frequency. Furthermore, strictly enforcing CPU core isolation (e.g., via Kubernetes QoS classes) does not isolate shared uncore resources. Consequently, uncore interference---where co-located workloads contend for memory bandwidth---can degrade performance unpredictably. In practice, to mitigate this risk, providers often default to maximum frequency (performance mode), sacrificing energy efficiency for reliability.

Several solutions have been proposed in the literature to address power–performance trade-offs, where  Reinforcement Learning (RL) has emerged as a promising approach for adaptive power control. Existing solutions often lack the granularity required for modern cloud environments. Most state-of-the-art RL approaches either control frequencies in isolation, optimize for utilization rather than Performance requirements, or neglect the impact of uncore interference.

To bridge this gap, we propose K8SPI (Kubernetes Power Irrigation), a Hierarchical Reinforcement Learning (HRL) controller that jointly optimizes core and uncore frequencies. K8SPI employs a two-stage control architecture to balance rapid performance requirement enforcement with energy minimization:

\begin{itemize}
    \item \textbf{Coarse-grained Restoration:} An agent that reacts to performance requirement violations by performing significant frequency adjustments to rapidly restore performance.
    \item \textbf{Fine-grained Refinement:} Once performance constraints are met, a second agent iteratively refines the configuration to minimize power consumption without triggering new violations.
\end{itemize}

To support this decision-making, K8SPI utilizes a multi-level telemetry stack:
\begin{itemize}
    \item \textbf{Low-level:} Per-microservice hardware performance counters capturing instruction execution and stall cycles.
    \item \textbf{Application-level:} Real-time microservice performance metrics (e.g., request latency).
    \item \textbf{Global-level:} Socket-wide counters reflecting memory controller and interconnect activity.
\end{itemize}

By synthesizing these signals, K8SPI effectively manages workload heterogeneity, dynamic load variations, and uncore interference.

We implement and evaluate K8SPI on a dedicated Kubernetes-based testbed that emulates cloud deployment conditions using benchmark-driven microservices. The evaluation covers three scenarios with increasing co-location and contention: (i) a single performance-sensitive microservice, (ii) a single performance-sensitive microservice co-located with best-effort workloads, and (iii) multiple performance-sensitive microservices sharing the same node under high contention.

Across all scenarios, K8SPI achieves a 23\%--30\% reduction in node-level power compared to the Linux performance governor operating at maximum frequency, while keeping latency violations below 2\%--3\%, even under dynamic load fluctuations and uncore interference. These results indicate that K8SPI can reduce power waste without compromising performance requirements in multi-tenant, cloud-native deployments.

In summary, the main contributions of this paper are as follows:
\begin{itemize}
    \item We propose K8SPI (Kubernetes Power Irrigation), a novel Hierarchical Reinforcement Learning (HRL) controller that jointly optimizes CPU core and uncore frequencies to reduce power consumption in cloud-native environments without violating microservice performance requirements.
    \item We develop an end-to-end framework for the rapid prototyping of RL-based power optimization policies, specifically tailored for latency-sensitive microservices.
    \item We implement K8SPI within a Kubernetes testbed, evaluating its robustness across increasingly complex scenarios, including isolated execution, best-effort co-location, and high-contention multi-instance deployments of latency-sensitive microservices.
    \item We demonstrate that K8SPI achieves a 23\%--30\% reduction in node-level power consumption compared to the default Linux performance governor operating at maximum frequency, while strictly maintaining performance requirement violations below 2\%--3\% under severe uncore interference and dynamic load fluctuations.
\end{itemize}

\subsection{Background \& Motivation}

\subsubsection{Latency-Sensitive Microservices}

Latency-sensitive microservices (e.g., real-time IoT analytics and interactive online services) operate under strict response-time Service Level Agreements (SLAs). In this work, latency is defined as the time elapsed between request arrival at the service and generation of the corresponding response. These constraints are strictly enforced because they directly impact user experience and contractual compliance.

Cloud services consume compute/memory resources (CPU, cache, memory bandwidth) in addition to storage/network; this work focuses on compute/memory only.

Microservices exhibit heterogeneous workload characteristics, which we categorize into three classes:
\begin{itemize}
    \item CPU-intensive (compute-dominated and sensitive to core frequency),
    \item memory-intensive (memory/bandwidth-dominated and typically more sensitive to uncore frequency), and
    \item mixed (joint compute and memory demands, sensitive to both core and uncore frequencies).
\end{itemize}

\subsubsection{Understanding Uncore Resource Contention}

Uncore resources, including the shared last-level cache (LLC) and memory bandwidth, are shared across CPU cores within a socket. This shared design introduces contention and can degrade performance when multiple memory-intensive microservices are co-located on the same node.

Uncore frequency plays a central role in this interference because it influences the operating point of the memory subsystem and, consequently, the achievable memory bandwidth at both per-core and socket-wide levels. Since the uncore is shared, bandwidth utilization is inherently interdependent: a core may achieve a high bandwidth in isolation at a given uncore frequency, but its effective throughput is bounded by the socket’s aggregate capacity and the concurrent demand from other cores.

For illustration, consider a 16-core server at a baseline uncore frequency with an aggregate socket bandwidth limit of 120~GB/s and an achievable single-core bandwidth of 15~GB/s. A single memory-intensive microservice pinned to one core can reach close to 15~GB/s. Under full co-location, if all 16 cores execute memory-bound workloads simultaneously, the aggregate demand exceeds 120~GB/s, and the effective bandwidth per core is reduced (e.g., to approximately 7.5~GB/s). This reduction increases memory stall time and can cause noticeable latency degradation for memory bandwidth-intensive microservices. Fig.~1 shows a 21\% per-core bandwidth drop under co-location.
\begin{figure}
\centering
    \includegraphics[width=0.3\textwidth]{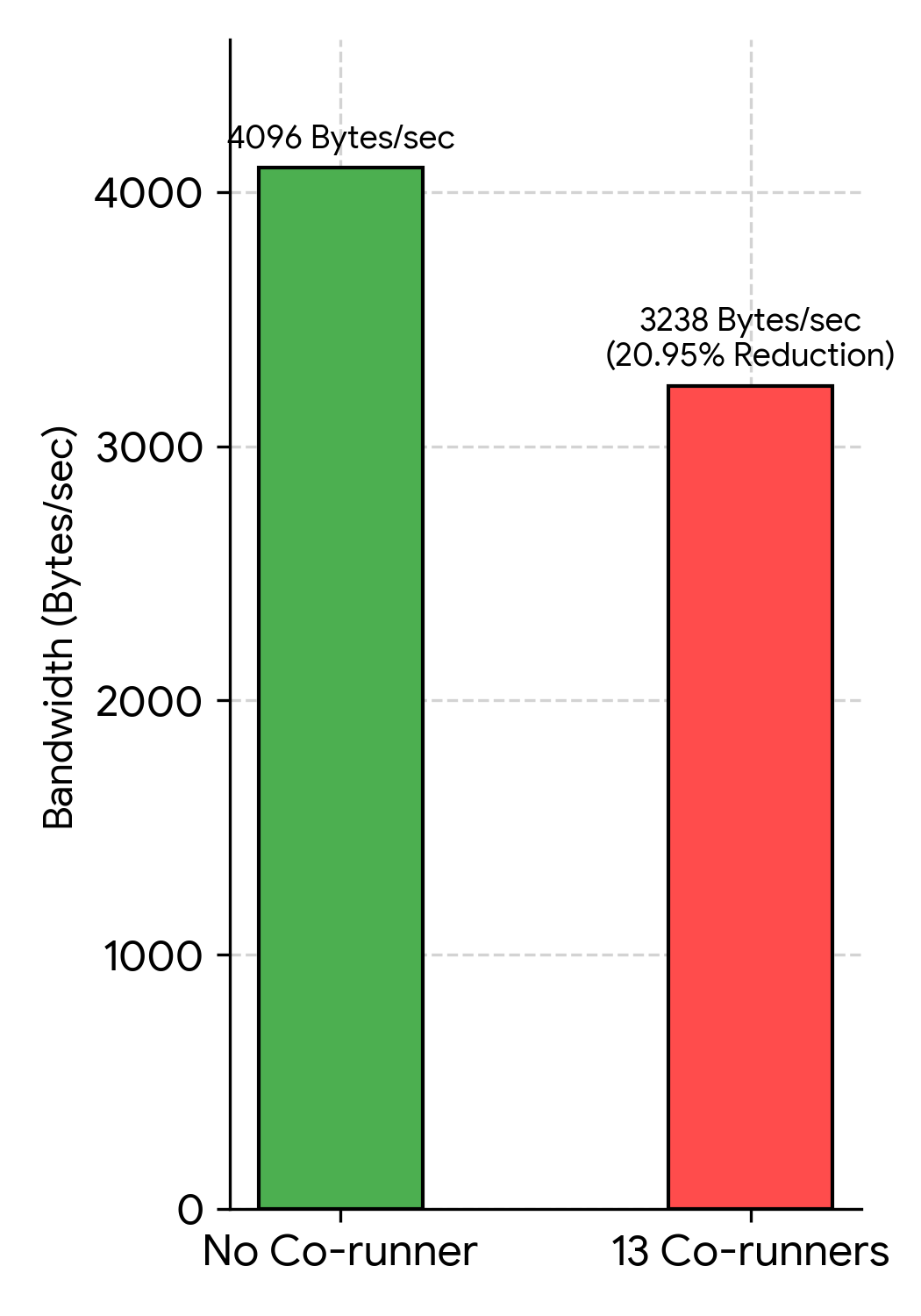}  
    \caption{Per-core memory bandwidth of a memory-intensive microservice running on a single CPU core in isolation versus co-located with 13 memory-intensive co-runners on a 14-core socket (core/uncore fixed at 2.6/2.4~GHz). Co-runners reduce per-core bandwidth by 21\%.}
    \label{fig:uncre_impact_L3}
\end{figure}

An efficient uncore frequency scaling policy that can scale up frequency under contention to improve bandwidth, and scale down when demand is low to avoid over-provisioning and reduce power.

\subsubsection{Frequency Management in Cloud Servers}

Cloud servers manage the performance--power trade-off using DVFS policies, commonly exposed as CPU frequency governors. These policies are typically categorized as static or dynamic.

Static governors (e.g., performance and powersave) operate at fixed operating points. The performance governor keeps frequencies at (or near) their maximum to prioritize performance, which helps performance-sensitive services satisfy latency requirements but often causes power waste through over-provisioning. In contrast, powersave prioritizes low frequencies to reduce power, which can degrade performance and increase the risk of performance requirement violations.

Dynamic governors (e.g., Intel P-state and Linux ondemand/schedutil) adapt frequencies online using utilization-driven heuristics derived from runtime signals. In general, core frequency is increased under higher compute demand and reduced when demand subsides.

\subsubsection{Core/Uncore Adjustment Policies}

\paragraph{Core frequency control}
Core DVFS policies typically adjust frequency using system-level indicators of compute demand, such as:  CPU load or IPC (or related hardware indicators), where higher instruction throughput can indicate compute-intensive execution and motivate higher frequency.

\paragraph{Uncore frequency governor}
While our solution targets Intel processors, the proposed methodology is architecture-agnostic and adaptable to other multi-core systems possessing, such as the AMD Zen architecture. The uncore is a substantial resource consumer, occupying roughly 30\% of the die area and driving approximately 20\% of total power consumption. Uncore frequency policies can be grouped into:
\begin{enumerate}
    \item CPU load-based governors: On some architectures (e.g., Intel Skylake and early Sapphire Rapids generations), uncore frequency is pushed to its maximum whenever any core reaches its base frequency. This approach ignores actual memory or cache demand. Such a strategy can lead to significant energy waste for CPU-bound applications where the application performance is insensitive to the memory latency. 
    \item Memory operation-based governors: More recent architectures (e.g., Intel Sierra Forest) introduce activity-driven uncore scaling, adjusting uncore frequency based on memory bandwidth usage and cache access patterns.
\end{enumerate}

\paragraph{Limitation: lack of application awareness}
Despite these advances, core and uncore governors are primarily driven by system-level signals and do not explicitly incorporate microservice performance requirements (e.g., latency). Consequently, they cannot reliably distinguish between (i) conservative scaling that preserves latency targets while reducing over-provisioning and (ii) aggressive downscaling that saves power but induces performance requirement violations, especially under heterogeneous workloads and multi-tenant interference.
\subsection{Reinforcement Learning}
Our work employs RL to develop a dynamic, performance-aware power optimization scheme. RL works by obtaining strategy improvements through continuous interactions with the changing environment in discrete time steps. At each step, an agent receives the current state and a reward (e.g., a function of the measured energy consumption and application-layer performance). It then chooses an action from the set of available actions (e.g., core and uncore configuration) and uses the action to configure the environment (e.g., the hardware). The environment will then move to a new state, and the reward associated with the transition is determined. The goal of the RL agent is to learn a policy that maximizes the expected cumulative reward, i.e., the overall power saving with guaranteed application performance. In this work, we choose to use a policy network optimized via Proximal Policy Optimization (PPO). Given the discrete nature of the action space and the need for fast online adaptation, PPO directly optimizes the policy and gracefully handles noisy environments. Unlike Q-learning-based methods, PPO utilizes incremental, continuous policy updates that ensure smooth adaptation to the evolving workloads and resource contention typical in dynamic cloud environments. This property makes the controller highly robust against runtime anomalies and short-term data drifts. Furthermore, PPO is generally less sensitive to hyperparameter tuning, providing the additional stability required for reliable online deployment.
\subsection{Related Work}

Prior work manages the power/energy--performance trade-off using DVFS and UFS. Early controllers are profile- or rule-based. RL has recently become a promising runtime alternative because it can adapt online under a dynamic environment.

\subsubsection{Non-RL Runtime Control}

A large body of prior work manages the power/energy--performance trade-off using DVFS and uncore frequency scaling (UFS) via profiling and rule-based runtime control. These methods are effective in structured settings (often HPC) where objectives are typically expressed as energy reduction under a bounded slowdown budget.

Cuttlefish~\cite{kumar2021cuttlefish} profiles MSRs to compute TIPI (memory requests per retired instruction), classifies workloads as compute- vs memory-bound, and triggers an exploration-based linear search to select core/uncore frequencies (CF/UF) minimizing joules per instruction (JPI). On a 20-core Intel Xeon Haswell E5-2650 with ten HPC benchmarks, it reports 19.4\% geometric-mean energy savings with 3.6\% slowdown versus the performance governor, but its deterministic classification policy can be brittle under phase changes and co-runner interference. DUFP~\cite{guermouche2022combining} combines package power capping with UFS using runtime FLOPS/s and bandwidth signals to detect phases and adjust RAPL caps within a tolerated slowdown (0--20\%), while tuning uncore using the same indicators; on NAS, HPL, and LAMMPS it reports improved energy efficiency, with strong results near a 10\% slowdown budget.

Complementary UFS-focused runtimes further validate uncore as a first-order control knob. DUF~\cite{andre2022duf} adapts uncore frequency online to reduce socket power/energy under bounded slowdown, and UPSCavenger~\cite{upscavenger2019sc} selects uncore frequency in a phase-aware manner for HPC workloads. Collectively, these efforts show that coordinated core--uncore tuning expands the energy--performance trade-off, but the objective is typically slowdown-centric rather than latency-SLO-centric.

For latency-critical services, non-RL controllers shift the objective from slowdown to explicit QoS preservation. PEGASUS~\cite{pegasus2014isca} uses feedback control for energy proportionality under service-level constraints; SleepScale~\cite{sleepscale2014isca} jointly selects DVFS and sleep states under QoS constraints; TimeTrader~\cite{timetrader2014} leverages tail-latency slack by slowing non-critical requests; and Gemini~\cite{gemini2020micro} uses learning-based prediction for per-query power/frequency decisions under tight latency targets.

\subsubsection{RL-Based DVFS/UFS Control Under QoS Constraints}

RL has emerged as a promising alternative for DVFS/UFS control under workload non-stationarity and co-location dynamics, because it can learn adaptive policies from telemetry rather than relying on fixed heuristics. In latency-sensitive, multi-workload environments, Twig~\cite{twig2020} applies deep RL with hardware counters to model QoS behavior and drive energy-efficient control under co-location. Hipster~\cite{hipster2017} combines RL with heuristic structure to manage latency-critical workloads alongside batch co-runners, using DVFS and runtime resource control to reduce energy while preserving QoS and improving throughput. Greeniac~\cite{greeniac2019} extends QoS-aware control to fog/edge clusters, framing frequency selection as a bandit problem under response-time constraints.

In service-centric networking, RL is also applied under explicit performance constraints. RL-ADR~\cite{wang2024rladr} selects frequency for a 5G user-plane function to minimize power under near-hard packet-drop constraints, using a policy library and conservative fallback to handle traffic distribution shifts; the authors report zero packet drops on long unseen traces while reducing power versus static baselines. GreenNFV~\cite{greennfv2023} uses actor--critic learning (DDPG) for NFV scheduling and tuning, including CPU frequency among multiple actuators to satisfy performance requirement/throughput targets while improving energy efficiency.

A second cluster shows the benefit of joint core--uncore control. Juan and Marculescu~\cite{juan2012islped} study reinforcement-based policies over core and uncore DVFS under power constraints, showing coordinated policies outperform core-only or uncore-only scaling under iso-power conditions. PManager~\cite{wang2021pmanager} co-optimizes power caps and UFS across phases using RL, and Gocht et al.~\cite{gocht2019hpcs} explore the core$\times$uncore space using Q-learning-style exploration with region awareness and direct energy feedback. A hierarchical RL design is also proposed in patent literature~\cite{zhu2024hierarchical}, but it lacks peer-reviewed evaluation. Finally, embedded platforms study RL-based DVFS under deadline/performance constraints (e.g., Q-learning in~\cite{biswas2017date} and DDQN with Linux CPUFreq in~\cite{zhou2023jsa}). Despite this progress, many studies still assume isolation or whole-socket occupancy. This abstraction does not match cloud nodes where multiple services share a socket and compete for uncore resources. Our objective is to satisfy per-service latency SLAs while minimizing power by dynamically scaling core and uncore frequencies under co-location, heterogeneous workload types, and time-varying load.
\section{K8SPI Architecture}

\subsection{System Overview}
\begin{figure}
\centering
    \includegraphics[width=0.8\textwidth]{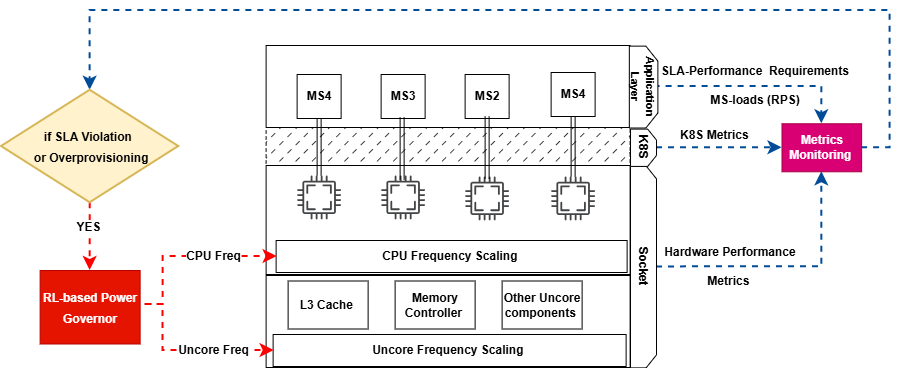}  
    \caption{K8SPI architecture overview.}
    \label{fig:K8SPI_architecture}
\end{figure}
In this paper, we propose K8SPI, an online, hierarchical reinforcement learning (HRL) based power governor tailored for multi-tenant cloud environments. Specifically, K8SPI targets heterogeneous microservices deployed via Kubernetes (K8s) under the Guaranteed Quality of Service (QoS) class, a configuration that allocates dedicated CPU cores to individual microservices. The primary objective of the K8SPI framework is to minimize the CPU socket power consumption without incurring performance degradation that violates the performance requirements of the hosted microservices.

As illustrated in Fig.~\ref{fig:K8SPI_architecture}, the K8SPI control loop consists of two primary modules: a multi-layered Metrics Monitoring module and an HRL-based Power Governor.

\subsection{Multi-Layered Telemetry and Monitoring}

To accurately capture the dynamic behavior of the system, the continuous monitoring module aggregates telemetry data across three distinct abstraction layers, providing both a global system view (i.e., socket-level view) and per-microservice visibility using:

\begin{itemize}
    \item \textbf{Application Layer:} Captures high-level performance metrics (e.g., target latency) and dynamic microservice loads, measured in requests per second (RPS) for each microservice.
    
    \item \textbf{Orchestration (K8s) Layer:} Gathers container- and pod-level metrics from the Kubernetes control plane to monitor deployment constraints and accurately map CPU cores to their respective microservices.
    
    \item \textbf{Hardware (Socket) Layer:} Collects lightweight, low-level hardware performance counter measurements (e.g., instructions per cycle, cache miss rates, and memory bandwidth utilization) at both the socket and core levels.
\end{itemize}

\subsection{Event-Triggered Hierarchical Power Governor}

K8SPI employs an event-driven control loop that reacts dynamically to the evaluated system state. The control loop actively evaluates the aggregated telemetry against current performance thresholds. As depicted by the decision logic in Fig.~X, the RL-based Power Governor is invoked exclusively when the monitoring subsystem detects an imminent performance requirement violation or a state of significant resource overprovisioning.

Upon identifying such conditions, the HRL-based Power Governor intervenes to concurrently modulate the dedicated CPU core frequency of the target microservice and the shared uncore frequency of the processor socket.

To efficiently navigate this expansive, multi-dimensional configuration space, the K8SPI governor relies on a two-stage hierarchical RL architecture:

\begin{itemize}
    \item \textbf{Stage 1 - Coarse Agent:} This agent makes large, rapid adjustments to the frequency settings. Its primary goal is to quickly mitigate performance violations or eliminate massive overprovisioning, bringing the system into a safe operational region. In doing so, it creates an optimal starting point for the subsequent tuning phase.
    
    \item \textbf{Stage 2 - Fine Agent:} Once the system is stabilized near the safe threshold, the fine agent takes over. It performs smaller, granular frequency refinements to minimize power consumption while ensuring that the performance requirement remains strictly satisfied.
\end{itemize}

\section{System Model and Problem Formulation}
\label{sec:system_model}

\subsection{Socket-Level Orchestration Architecture}

We consider a multi-tenant cloud compute node comprising multiple physical processor sockets; K8SPI acts at the socket level. The socket encompasses a set of physical CPU cores, denoted by $\mathcal{C}$, and hosts a set of heterogeneous, latency-critical microservices, $\mathcal{M}$. To guarantee strict compute isolation and satisfy Kubernetes Guaranteed Quality of Service (QoS) requirements, we assume an exclusive deployment mapping where each microservice $m_i \in \mathcal{M}$ is pinned to a dedicated core $c_i \in \mathcal{C}$.

The hardware exposes two primary control domains for dynamic voltage and frequency scaling (DVFS), and uncore frequency scaling (UFS). First, each core $c_i$ operates at an independent core frequency, $f_{core,i}$, bounded by $[f_{core}^{min}, f_{core}^{max}]$. Second, all cores within the socket share a common uncore subsystem, which includes the Last Level Cache (LLC) and memory controllers. This shared domain operates at a global uncore frequency, $f_{uncore}$, bounded by $[f_{uncore}^{min}, f_{uncore}^{max}]$.

\subsection{Workload and Performance Model}
The system operates under an event-triggered control paradigm. Let $t_k$ denote the timestamp of the $k$-th control invocation epoch, triggered by either an imminent performance violation or a state of massive resource overprovisioning. 

Each microservice $m_i$ is characterized by an architectural workload profile, $\omega_i$ (e.g., compute-bound or memory-bound), and experiences a dynamic incoming request load, $\lambda_i(t_k) \in [0, 1]$. The performance of $m_i$ is strictly governed by a target latency constraint, $L_{target,i}$. 

The measured tail latency at epoch $t_k$, denoted as $L_{current,i}(t_k)$, is modeled as the sum of its isolated execution latency and a contention time penalty:

\begin{equation}
\label{eq:latency_model}
L_{current,i}(t_k) = \Phi \left( f_{core,i}(t_k), f_{uncore}(t_k), \lambda_i(t_k), \omega_i \right) + \Delta L_{interf, i}(t_k)
\end{equation}

where $\Phi(\cdot)$ represents the baseline latency function of the microservice executed without noisy neighbors. The term $\Delta L_{interf, i}(t_k) \geq 0$ encapsulates the latency time penalty induced by uncore interference. Because latency measures execution delay, interference strictly adds to this delay. This degradation occurs because co-located microservices aggressively compete for the shared memory bandwidth. While increasing $f_{uncore}$ expands the aggregate socket bandwidth, the effective throughput available to $m_i$ decreases as co-runner demand increases, causing memory stall cycles that strictly inflate $L_{current,i}$.

Crucially, because $\Phi(\cdot)$ and $\Delta L_{interf, i}$ depend on highly dynamic runtime behaviors, opaque microarchitectural state interactions (e.g., core pipeline stalls and cache hierarchy thrashing), and complex, non-linear resource contention across the entire socket, deriving an exact analytical model is computationally intractable. Instead, these highly dynamic and coupled interactions must be approximated and learned iteratively at runtime.

\subsection{Power Model}
The total power consumption of the socket, $P_{socket}(t_k)$, is decomposed into the dynamic power of the active cores, the dynamic power of the shared uncore domain, and the baseline static leakage power ($P_{static}$):

\begin{equation}
\label{eq:power_model}
P_{socket}(t_k) = \sum_{i \in \mathcal{C}} P_{dyn}^{core}(f_{core,i}(t_k), \lambda_i(t_k)) + P_{dyn}^{uncore}(f_{uncore}(t_k), \vec{\lambda}(t_k)) + P_{static}
\end{equation}

Because dynamic power scales proportionally with the switching frequency and the square of the supply voltage ($P \propto V^2f$), aggressive scaling of both core and uncore frequencies provides the primary lever for power minimization. Crucially, in modern high-density processors, the uncore subsystem represents a massive physical domain. Consequently, $P_{dyn}^{uncore}$ consumes a disproportionately dominant fraction of the total socket power compared to any individual core's dynamic power ($P_{dyn}^{core}$). Therefore, the K8SPI controller treats the uncore frequency not just as a performance knob, but as the most critical variable for achieving node-wide energy efficiency, albeit one highly coupled to the aggregate memory transaction rate $\vec{\lambda}(t_k)$ generated by all cores on the socket.

Similar to the latency model, explicitly formalizing the complex, non-linear coupling between highly dynamic multi-core workloads, joint frequency scaling, and holistic socket power consumption across diverse hardware is practically infeasible at runtime. Therefore, the exact power-performance manifold is treated as a highly dynamic, black-box environment whose underlying transition dynamics will be implicitly learned via the Deep Reinforcement Learning policy.

\subsection{Problem Definition}
The objective of the proposed K8SPI framework is to determine the optimal frequency scaling actions at each event epoch $t_k$ to minimize the total socket power consumption, while strictly guaranteeing that no microservice violates its performance requirement. The constrained optimization problem is formulated as:

\begin{equation}
\label{eq:objective}
\min_{ \vec{f}_{core}, f_{uncore} } \sum_{k=0}^{K} P_{socket}(t_k)
\end{equation}

\begin{equation}
\label{eq:constraint}
\text{subject to} \quad L_{current,i}(t_k) \leq L_{target,i} \quad \forall m_i \in \mathcal{M}, \forall k
\end{equation}
\begin{equation}
\label{eq:bounds}
f_{core}^{min} \leq f_{core,i}(t_k) \leq f_{core}^{max}, \quad f_{uncore}^{min} \leq f_{uncore}(t_k) \leq f_{uncore}^{max}
\end{equation}

This formulation highlights the core control challenge: minimizing Equation \ref{eq:power_model} by lowering frequencies directly penalizes the latency function in Equation \ref{eq:latency_model}, thereby risking violations of the strict constraint in Equation \ref{eq:constraint} amid unpredictable uncore interference.

Because the functions governing power (Eq. \ref{eq:power_model}) and latency (Eq. \ref{eq:latency_model}) exhibit complex, non-linear, and highly dynamic behaviors at runtime across both core and uncore domains, traditional convex optimization and static heuristic control are inadequate. Consequently, we frame this optimization as a Markov Decision Process (MDP). K8SPI leverages a Deep Reinforcement Learning agent to map real-time telemetry states to optimal frequency actions, inherently learning a robust control policy capable of navigating the complex trade-offs between strict performance requirement enforcement and maximum energy efficiency.

\section{Hierarchical Reinforcement Learning Controller}
\label{sec:hrl_controller}

To navigate the high-dimensional joint frequency space efficiently, we propose a Hierarchical Reinforcement Learning (HRL) architecture. The hierarchical approach decomposes the DVFS adjustment into two sequential decisions. 

By splitting the action into coarse and fine stages, the hierarchical controller reduces the search space and explicitly separates the objectives of rapid SLA stabilization and fine-grained energy minimization. Triggered whenever a performance requirement violation or massive resource overprovisioning crosses a defined threshold, the control loop operates over a fixed two-step episode: Agent A1 (coarse) selects a large frequency delta to immediately exit the performance requirement violation or reduce gross overprovisioning, providing a safe and feasible starting point. Agent A2 (fine) then refines A1's action by taking smaller delta steps to safely maximize energy savings before the episode terminates.

\subsection{State Representation and Observations}
At each decision step, the active agent receives a structured observation representing the system's telemetry. The state space is composed of three primary components:

\begin{itemize}
    \item \textbf{Service-level vector ($x_s$):} Contains microservice-specific performance counters at the hardware level (e.g., IPC, bandwidth per core) and at the application level (e.g., current latency). This vector is primarily used to guide the core frequency scaling decisions.
    \item \textbf{Socket-level vector ($x_u$):} Contains aggregated socket-wide counters, reflecting the "outside-the-core" view and the resource pressure exerted by concurrent, co-located microservices. This vector is primarily used to guide uncore frequency scaling.
    \item \textbf{Action Mask ($m$):} A mask that marks infeasible actions as invalid to prevent the agent from choosing frequencies outside the allowable hardware ranges.
\end{itemize}

The specific metrics tracked in the observation space are detailed in Table \ref{tab:observation_metrics}. These metrics are highly curated as they are the primary factors determining the hardware-level performance of the microservice. Crucially, they allow the RL agent to identify the architectural bottleneck (e.g., compute-bound vs. memory-bound) at runtime.

\begin{table}[htbp]
\centering
\caption{RL Agent Observation Space Metrics}
\label{tab:observation_metrics}
\renewcommand{\arraystretch}{1.2}
\begin{tabular}{@{}ll@{}}
\toprule
\textbf{Metric Name} & \textbf{Description} \\ 
\midrule
\multicolumn{2}{c}{\textbf{Service-Level Metrics (Used to guide core)}} \\ 
\midrule
Core\_L3\_Misses & L3 cache misses on the service core \\
Core\_L3\_Hits & L3 cache hits on the service core \\
Core\_Mem\_Bandwidth & Local memory bandwidth used by the service \\
Service\_Avg\_Latency & Average latency of service requests \\
Service\_Target\_Latency & Target latency of service requests \\
Core\_MPKC & Misses per thousand instructions \\
Core\_IPC & Instructions per cycle \\
Core\_KIPS & Kilo Instructions per Second \\
CPU\_Time\_System & Time spent in system/kernel mode \\
CPU\_Time\_User & Time spent in user mode \\
CPU\_Utilization & Container's total CPU usage (\%) \\
Core\_Frequency & Core frequency assigned to the service \\
Core\_L3\_Occupancy & L3 cache usage by the service core \\
\midrule
\multicolumn{2}{c}{\textbf{Socket-Level Metrics (Used to guide uncore)}} \\ 
\midrule
Socket\_L3\_Misses & Aggregate L3 cache misses on the socket \\
Socket\_L3\_Hits & Aggregate L3 cache hits on the socket \\
Socket\_Mem\_Bandwidth & Total memory bandwidth on the socket \\
Socket\_L3\_Occupancy & Overall L3 cache occupancy \\
Socket\_Power\_CPU & Package-level CPU power consumption (W) \\
Socket\_Power\_DRAM & DRAM power consumption (W) \\
Socket\_Uncore\_Frequency & Current uncore frequency on the socket \\
\bottomrule
\end{tabular}
\end{table}

\subsubsection{Frequency-Conditioned State Normalization}
\label{sec:state_normalization}

Standard global normalization fails to capture how physical hardware boundaries shift under Dynamic and Frequency Scaling. To ensure stable convergence and explicitly embed domain knowledge, we introduce a frequency-conditioned online normalization strategy. The environment maintains independent running statistics (mean $\mu$, standard deviation $\sigma$) for each unique core and uncore frequency pair $(f_{\text{core}}, f_{\text{uncore}})$. The telemetry vector $x$ is normalized via an online z-score strictly within its active frequency bucket:
\begin{equation}
x^{(\text{norm})} = \frac{x - \mu_{(f_{\text{core}}, f_{\text{uncore}})}}{\sigma_{(f_{\text{core}}, f_{\text{uncore}})} + \xi}
\end{equation}

This bucketed approach enables the RL agent to intelligently isolate architectural bottlenecks. If compute-related metrics (e.g., KIPS) yield high positive z-scores during an performance requirement violation, the agent infers core saturation and scales up $f_{\text{core}}$. Conversely, saturating memory metrics (e.g., bandwidth) prompts the agent to scale $f_{\text{uncore}}$. While this accurately captures regime-specific hardware limits, it introduces a minor cold-start phase where newly visited frequency pairs require a brief initialization window to stabilize their statistics.

\subsection{Hierarchical Action Space}
Rather than exploring a massive flat joint action space (e.g., 289 actions), the two-agent design drastically reduces the dimensionality.
\begin{itemize}
    \item \textbf{Coarse Stage (Agent A1):} Selects broad adjustments with $\Delta f \in \{-0.8, 0, +0.8\}$\,GHz for both the core and uncore knobs, yielding a highly focused space of $3 \times 3 = 9$ actions.
    \item \textbf{Fine Stage (Agent A2):} Refines the coarse adjustment by selecting smaller deltas with $\Delta f \in \{-0.4, -0.2, 0, +0.2, +0.4\}$\,GHz, yielding $25$ possible actions.
\end{itemize}

For both domains, the selected action is a discrete frequency delta applied to the current operating state:
\begin{equation}
f_{\text{core}}^{(new)} = f_{\text{core}} + \Delta f_{\text{core}}, \quad f_{\text{uncore}}^{(new)} = f_{\text{uncore}} + \Delta f_{\text{uncore}}
\end{equation}

To strictly prevent the agent from selecting out-of-bound frequencies, the environment applies a dynamic binary mask $\mathcal{M}$ that filters infeasible deltas prior to action selection:
\begin{equation}
\mathcal{M}(\Delta f_{\text{core}}, \Delta f_{\text{uncore}}) = 
\begin{cases} 
1, & \text{if } f_{\text{core}}^{\min} \leq f_{\text{core}} + \Delta f_{\text{core}} \leq f_{\text{core}}^{\max} \text{ \textbf{and} } f_{\text{uncore}}^{\min} \leq f_{\text{uncore}} + \Delta f_{\text{uncore}} \leq f_{\text{uncore}}^{\max} \\
0, & \text{otherwise}
\end{cases}
\end{equation}

\subsection{performance Gap Quantization}
To stabilize the learning process and prevent control jitter, the performance state of each microservice is evaluated via a quantized latency gap. First, the raw relative latency gap for a given microservice is defined as:
\begin{equation}
\label{eq:raw_gap}
g_{raw} = \frac{L - L^*}{L^*}
\end{equation}
where $L$ is the current measured average latency and $L^*$ is the strict target latency constraint. Under this formulation, a positive value indicates an performance requirement violation, whereas a negative value indicates resource overprovisioning. 

Crucially, the magnitude of this raw gap serves as the primary trigger for corrective action. The control loop is only invoked when the performance deviates beyond an acceptable boundary—for instance, an acceptable latency degradation threshold (e.g., $g_{raw} > +15\%$) or a severe overprovisioning threshold (e.g., $g_{raw} < -15\%$).

Rather than feeding the continuous, noisy $g_{raw}$ directly to the RL agent, the environment maps it to a discrete, bucketed gap $g(L, f_c, f_u)$ using a three-tiered logic:
\begin{equation}
\label{eq:quantized_gap}
g(L, f_c, f_u) = 
\begin{cases} 
0, & \text{if hardware-saturated} \\
0, & \text{if } |g_{raw}| < \epsilon \\
\text{sgn}(g_{raw}) \delta \lceil |g_{raw}| / \delta \rceil, & \text{otherwise}
\end{cases}
\end{equation}
This quantization strictly enforces three operational rules to ensure predictable power transitions:
\begin{enumerate}
    \item \textbf{Deadzone ($\epsilon$):} A tight tolerance band (e.g., $\tau = 5\%$) masks small, transient latency fluctuations. This prevents jitter and unnecessary frequency toggling when the system is operating near the target.
    \item \textbf{Step-Based Scaling ($\delta$):} The gap is discretized into fixed intervals (e.g., $\delta = 10\%$) to group state observations into distinct buckets, rendering the optimization landscape more stable for the agent.
    \item \textbf{Hardware-Aware Safety Cap:} The gap is forced to zero if the system is hardware-saturated. Specifically, this occurs if the performance requirement is satisfied but frequencies are already at their lowest supported bounds ($L \leq L^* \land f_c = f_{c,min} \land f_u = f_{u,min}$), or if the performance requirement is violated but frequencies are already maxed out ($L \geq L^* \land f_c = f_{c,max} \land f_u = f_{u,max}$). This hardware awareness stops the controller from attempting physically impossible adjustments.
\end{enumerate}

\subsection{Hierarchical Reward and Credit Assignment}
\label{sec:hier_reward}

The reward signals are engineered to encode both strict performance requirement constraint satisfaction and power efficiency. The hierarchical controller operates over a fixed two-step episode:
\begin{equation}
s_0 \xrightarrow[\Delta f_c^{(1)}, \Delta f_u^{(1)}]{\text{A1 (coarse)}} s_1 \xrightarrow[\Delta f_c^{(2)}, \Delta f_u^{(2)}]{\text{A2 (fine)}} s_2
\end{equation}
where Agent A1 performs a coarse move transitioning the system to intermediate state $s_1$, and Agent A2 refines it to reach the terminal state $s_2$. Let $g_0$, $g_1$, and $g_2$ represent the quantized performance gaps at the pre-A1, post-A1, and terminal stages, respectively.

\paragraph{Universal Performance Potential}
To evaluate performance requirement compliance consistently across both hierarchical stages, we define a universal performance potential function $\Phi(g)$ based on the quantized gap:
\begin{equation}
\Phi(g) = 
\begin{cases}
-(2 + g), & g > \epsilon \quad \text{(violation)} \\[0.5em]
1 - |g|, & g \leq \epsilon \quad \text{(safe)}
\end{cases}
\end{equation}
where $\epsilon$ represents the SLA safety threshold (e.g., $\epsilon = 0.05$). This piecewise function heavily penalizes performance requirement relatively to variation magnitude (g)  while rewarding proximity to the target latency when operating in the safe zone.   

\paragraph{Etiquette Penalty}
To embed domain knowledge and prevent unsafe throttling, an etiquette penalty applies at both steps $t \in \{1, 2\}$. It strictly penalizes the active agent for decreasing both frequencies if the system is already in a state of performance requirement violation:
\begin{equation}
\psi_t = \eta_{\text{both}} \cdot \mathbb{I}\left[g_{t-1} > 0 \land \Delta f_c^{(t)} < 0 \land \Delta f_u^{(t)} < 0\right] \qquad (\text{with } \eta_{\text{both}} = 0.5)
\end{equation}

\subsection{Agent A2 (Fine) Reward Function}
The goal of the fine agent is to refine the frequencies to minimize power once the system is near or within the safe region. Its total reward, $R_{A2}$, evaluates the terminal performance state $g_2$ and adds an efficiency bonus $R^*$ if the SLA is satisfied, minus its step-specific etiquette penalty:
\begin{equation}
R_{A2} = \Phi(g_2) + \mathbb{I}[g_2 \leq \epsilon] \cdot R^* - \psi_2
\end{equation}

To explicitly favor low-power operating points, each discrete hardware frequency is mapped to a normalized index $a_x \in [0, 2]$, where $0$ is the minimum supported frequency and $2$ is the maximum. The efficiency bonus is defined as:
\begin{equation}
R^* = w_c(1 - a_{c,2}) + w_u(1 - a_{u,2})
\end{equation}

When the performance requirement is met ($g_2 \leq \epsilon$), the agent balances two competing objectives. The performance term $\Phi(g_2) = 1 - |g_2|$ encourages the agent to operate closely to the target latency, minimizing unnecessary performance slack and overproviding. Concurrently, the efficiency bonus $R^*$ drives power reduction. Because dynamic power scales directly with operating frequency, core and uncore frequencies serve as reliable proxies for power consumption. The normalization mapping to $a_x \in [0, 2]$ ensures that selecting the minimum supported frequency yields a maximum positive reward (e.g., $1 - 0 = 1$), whereas operating at the maximum frequency introduces a penalty (e.g., $1 - 2 = -1$). Furthermore, we configure the weights such that $w_c < w_u$; consequently, raising the core frequency is less penalized than raising the uncore frequency, accurately reflecting the higher power cost of the socket-wide uncore domain.

\subsection{Agent A1 (Coarse) Reward Function}
The primary goal of the coarse agent is to rapidly reduce the latency gap and create an optimal, safe starting point for A2's fine-tuning. Because A1 transitions the system to the intermediate state $s_1$, its immediate performance reward is $\Phi(g_1)$. 

To ensure cohesive teamwork and prevent A1 from acting greedily, A1's total reward directly incorporates A2's terminal outcome, while subtracting its own step-specific etiquette penalty ($\psi_1$):
\begin{equation}
R_{A1} = \Phi(g_1) - \psi_1 + R_{A2}
\end{equation}
This elegantly coupled design forces A1 to select coarse actions that not only immediately improve the SLA state ($\Phi(g_1)$) but also position the intermediate state such that A2 can successfully minimize energy without triggering a terminal violation ($R_{A2}$).

\subsection{Runtime Orchestration and Action Aggregation}
\label{sec:runtime_orchestration}

In a live production environment, K8SPI operates as an event-triggered control loop monitoring the performance gaps of all co-located microservices on a shared socket. When a specific microservice $m_i \in \mathcal{M}$ actively violates its performance requirement/severely over-provisioned ($g_i > +/-20\%$), the RL controller proposes a corrective joint frequency action $(f_{\text{core}, i}, f_{\text{uncore}, i})$ per microservice. However, the physical hardware imposes distinct granularity constraints for applying these independent per-service actions.

When come to the proposed core frequency is applied directly and independently to the specific core hosting $m_i$ (as microservices are pinned to a single cpu core):
\begin{equation}
f_{\text{core}}(c_i) = f_{\text{core}, i}
\end{equation}

Conversely, the uncore subsystem is a shared, socket-wide resource. To resolve conflicting uncore frequency proposals among co-located services while ensuring the performance requirements of all microservices are met, the final uncore frequency $f_{\text{uncore}}^*$ applied to the socket is the maximum of all individual proposals:
\begin{equation}
f_{\text{uncore}}^* = \max_{m_i \in \mathcal{M}} \{ f_{\text{uncore}, i} \}
\end{equation}

This pessimistic aggregation guarantees that the most memory-constrained service is strictly protected against uncore throttling, while node-wide power savings are successfully captured only when all co-located services mutually tolerate a scaled-down uncore frequency.

\section{Scalable Implementation and Training Methodology}
\label{sec:implementation}
\begin{figure}[b]
\centering
    \includegraphics[width=0.8\textwidth]{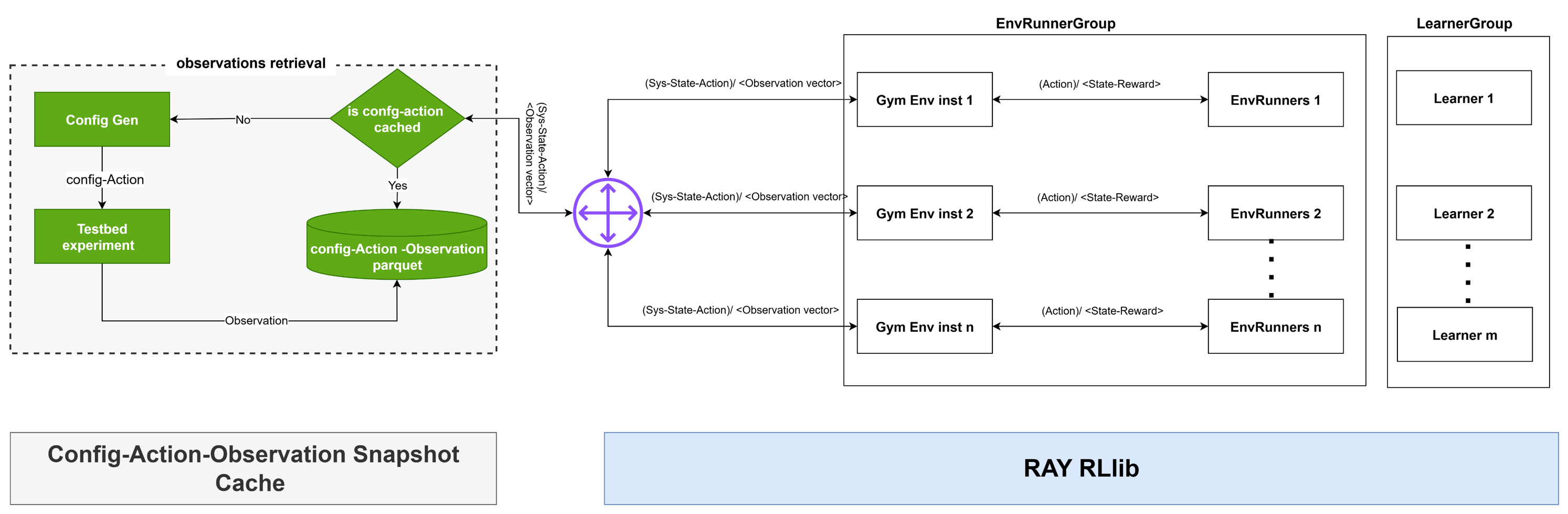}  
    \caption{Fast RL Prototyping Framework Architecture Overview.}
    \label{fig:RLFAST}
\end{figure}
Hardware-in-the-loop RL is slowed by multi-second feedback: each frequency action must be applied and then observed through the monitoring stack (e.g., Prometheus) before the next step. Since RL needs thousands of trials, this makes training and iterative tuning (reward design, hyperparameters) prohibitively slow on a live Kubernetes node.

\subsection{Distributed Training and Snapshot Caching}
\label{sec:distributed_training}

To accelerate learning, K8SPI uses Ray RLlib to enable distributed training with a centralized learner coordinating multiple concurrent environment runners. However, running many environments is costly because each environment ultimately requires execution on the physical node (frequency actions are hardware operations and cannot be virtualized). K8SPI resolves this by inserting an offline Config--Action--Observation Snapshot Cache. Fig.~\ref{fig:RLFAST} shows the architecture of the fast RL prototyping framework. When an RL agent instance encounters a new $\langle \text{state}, \text{action} \rangle$ pair, the framework performs a one-time hardware profiling run: it deploys the state (the specific microservice executing its corresponding benchmark under a defined uncore stress level), applies the selected frequency configuration on the testbed, and collects telemetry over a sustained 5-minute window to average out transient effects. The resulting high-fidelity observation vector is stored in a Parquet-backed datastore. Future visits to the same $\langle \text{state}, \text{action} \rangle$ pair are served instantly from the cache, enabling safe parallel RLlib training without live-hardware latency.

\section{Experimental Methodology}
\label{sec:experimental_setup}
\begin{figure}[t]
\centering
    \includegraphics[width=0.8\textwidth]{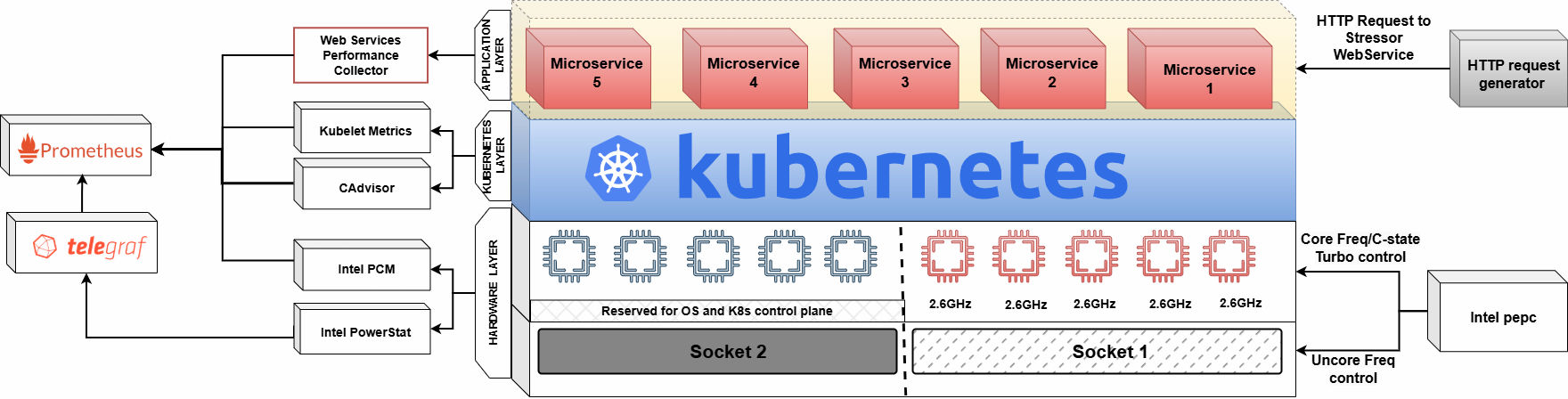}  
    \caption{Testbed Architecture Overview.}
    \label{fig:Testbed}
\end{figure}
\subsection{Testbed Infrastructure and Software Stack}
\label{sec:testbed}

\begin{table}[!h]
\centering
\caption{Testbed Hardware and Software Configuration}
\label{tab:testbed_configuration}
\setlength{\tabcolsep}{4pt}
\renewcommand{\arraystretch}{1.1}
\footnotesize
\begin{tabular}{@{} l l l l @{}}
\toprule
\textbf{Attribute} & \textbf{Specification} & \textbf{Attribute} & \textbf{Specification} \\
\midrule
CPU Model          & Intel Xeon Gold 6132  & Threads/Core       & 1 (HT Disabled) \\
Cores/Socket       & 14                    & Sockets            & 2 \\
NUMA Nodes         & 2                     & Core Freq (GHz)    & 1.0 -- 2.6 \\
L3 Cache           & 38.5 MB/socket        & Uncore Freq (GHz)  & 1.2 -- 2.4 \\
OS                 & Ubuntu 22.04.5        & Kubernetes QoS     & Guaranteed \\
Turbo \& C-States  & Disabled              & Memory Swap        & Disabled \\
Core Governor      & \texttt{userspace}    & Uncore Governor    & \texttt{userspace}  \\
\bottomrule
\end{tabular}
\vspace{-0.3cm}
\end{table}

To ensure stable, reproducible, and noise-free Reinforcement Learning observations and power measurements, we constructed a strictly controlled, dual-socket Kubernetes testbed wchich it congration illustrated in Table~\ref{tab:testbed_configuration} and in Fig~\ref{fig:Testbed}. 

To prevent background OS noise from corrupting the RL agent's telemetry or power measurements, we enforced strict hardware-level isolation. Socket 1 was dedicated to executing the experimental microservices and the telemetry stack. Socket 2 was reserved entirely for the Kubernetes control plane and operating system background processes. Moreover, to eliminate unpredictable power fluctuations, several BIOS and kernel-level features were explicitly disabled (see Table~\ref{tab:testbed_configuration}). The telemetry stack utilized Intel PCM for hardware performance counters, Intel PowerStat (i.e., Intel RAPL) for socket-level ground-truth power measurements, and cAdvisor alongside Kubelet metrics for per-microservice resource usage and core-pinning resolution. Finally, we employed custom scripts to measure the end-to-end latency per HTTP request for each hosted service, establishing the application-level performance of each microservice. Data was scraped and aggregated using a telemetry stack comprising Telegraf and Prometheus. To prevent load-generation overhead from interfering with the testbed, the traffic generators were hosted remotely and triggered the microservices via HTTP. To control the core and uncore frequencies during the experiments, we utilized the open-source Intel \texttt{pepc} library, which provides a sophisticated power control toolkit.

\subsection{Benchmarks and Workloads}
\label{sec:benchmarks}

We construct a benchmark-driven microservice corpus using two micro-benchmark suites: Stress-NG (primarily CPU-bound and mixed behaviors) and Parallel Memory Bandwidth (PMBW) (memory-bound behaviors with strong sensitivity to the shared memory subsystem and uncore frequency). Each selected benchmark is containerized and deployed as a stateless HTTP microservice with fixed resource allocations (1 CPU and 2 GB memory). Each request triggers a bounded unit of benchmark work executed at a controlled load level. For each benchmark, we define 4–5 discrete load settings (e.g., a fixed number of memory-copy operations) to expose different demand workload levels.
\begin{figure}[t]
\centering
    \includegraphics[width=0.7\textwidth]{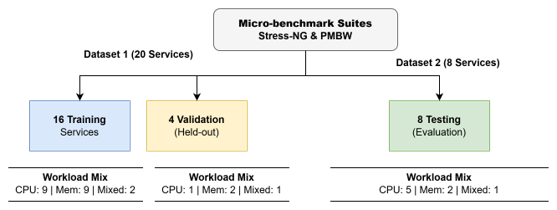}  
    \caption{Micro-benchmark Dataset Split for Model Development and Evaluation.}
    \label{fig:dataset}
\end{figure}
To separate learning and evaluation from final testing, we create two disjoint datasets. Dataset 1 contains 20 services, split into 16 for training and 4 held out for RL model evaluation. Dataset 2 contains 8 additional services reserved exclusively for final testing and validation. Fig.~\ref{fig:dataset} illustrates the dataset partitioning across the training, evaluation, and test phases. 

\begin{figure*}[h]
    \centering
    
    \begin{subfigure}[t]{0.24\textwidth}
        \centering
        \includegraphics[width=\linewidth]{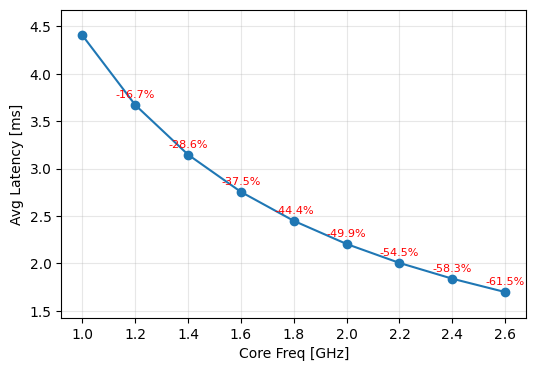}
        \caption{Latency vs. Core Freq.}
        \label{fig:ackermann_lat_core}
    \end{subfigure}\hfill
    \begin{subfigure}[t]{0.24\textwidth}
        \centering
        \includegraphics[width=\linewidth]{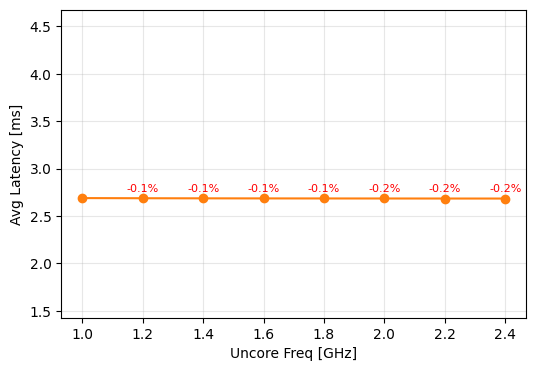}
        \caption{Latency vs. Uncore Freq.}
        \label{fig:ackermann_lat_uncore}
    \end{subfigure}\hfill
    \begin{subfigure}[t]{0.24\textwidth}
        \centering
        \includegraphics[width=\linewidth]{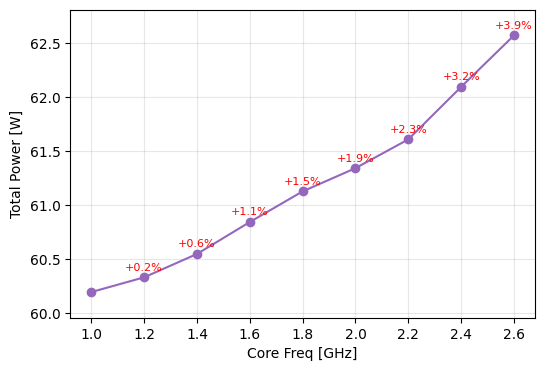}
        \caption{Power vs. Core Freq.}
        \label{fig:ackermann_pow_core}
    \end{subfigure}\hfill
    \begin{subfigure}[t]{0.24\textwidth}
        \centering
        \includegraphics[width=\linewidth]{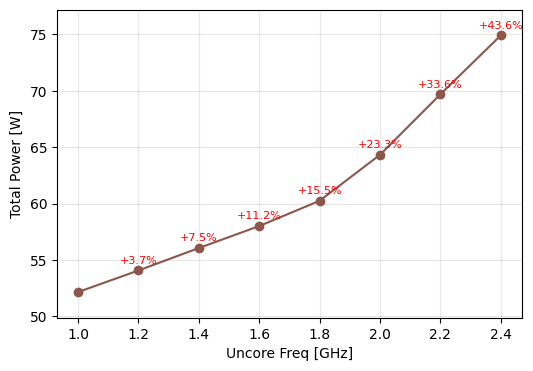}
        \caption{Power vs. Uncore Freq.}
        \label{fig:ackermann_pow_uncore}
    \end{subfigure}

    \caption{Impact of core and uncore frequency scaling on the CPU-bound \texttt{BG\_CPUAckermann} microservice.}
    \label{fig:ackermann_core_uncore_impact}
\end{figure*}

\begin{figure*}[h]
    \centering
    
    \begin{subfigure}[t]{0.24\textwidth}
        \centering
        \includegraphics[width=\linewidth]{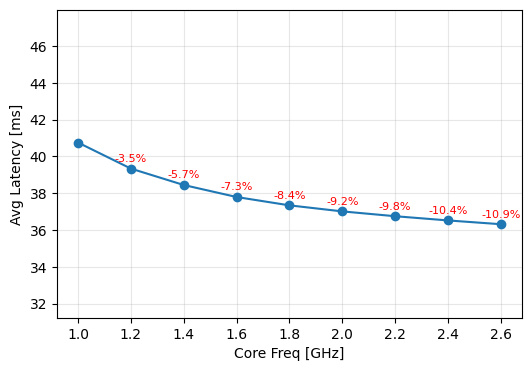}
        \caption{Latency vs. Core Freq.}
        \label{fig:memrateflush_lat_core}
    \end{subfigure}\hfill
    \begin{subfigure}[t]{0.24\textwidth}
        \centering
        \includegraphics[width=\linewidth]{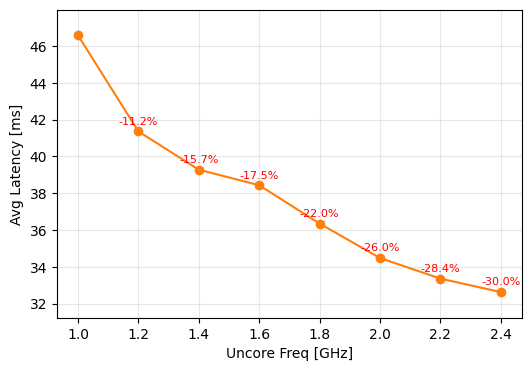}
        \caption{Latency vs. Uncore Freq.}
        \label{fig:memrateflush_lat_uncore}
    \end{subfigure}\hfill
    \begin{subfigure}[t]{0.24\textwidth}
        \centering
        \includegraphics[width=\linewidth]{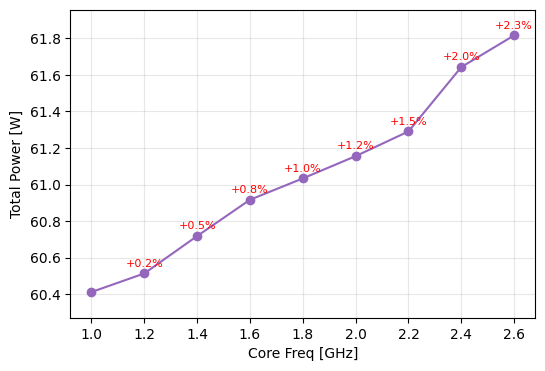}
        \caption{Power vs. Core Freq.}
        \label{fig:memrateflush_pow_core}
    \end{subfigure}\hfill
    \begin{subfigure}[t]{0.24\textwidth}
        \centering
        \includegraphics[width=\linewidth]{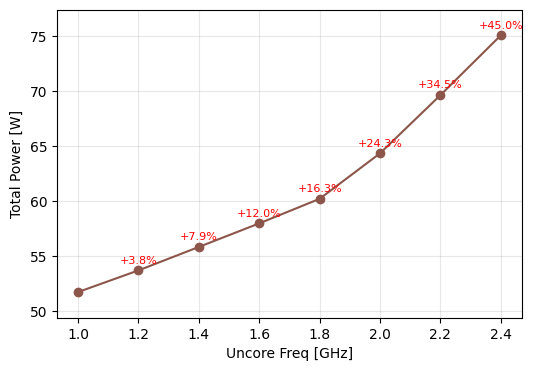}
        \caption{Power vs. Uncore Freq.}
        \label{fig:memrateflush_pow_uncore}
    \end{subfigure}

    \caption{Impact of core and uncore frequency scaling on the memory-bound \texttt{BG\_MemrateFlush} microservice.}
    \label{fig:MemrateFlush_core_uncore_impact}
\end{figure*}

\begin{figure*}[h]
    \centering
    
    \begin{subfigure}[t]{0.24\textwidth}
        \centering
        \includegraphics[width=\linewidth]{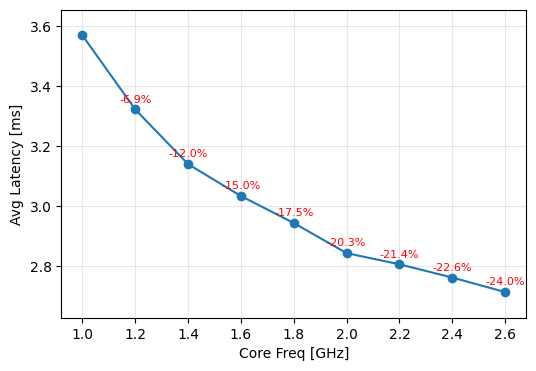}
        \caption{Latency vs. Core Freq.}
        \label{fig:memthrash_lat_core}
    \end{subfigure}\hfill
    \begin{subfigure}[t]{0.24\textwidth}
        \centering
        \includegraphics[width=\linewidth]{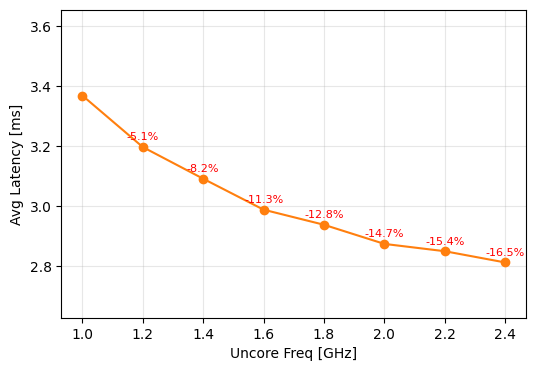}
        \caption{Latency vs. Uncore Freq.}
        \label{fig:memthrash_lat_uncore}
    \end{subfigure}\hfill
    \begin{subfigure}[t]{0.24\textwidth}
        \centering
        \includegraphics[width=\linewidth]{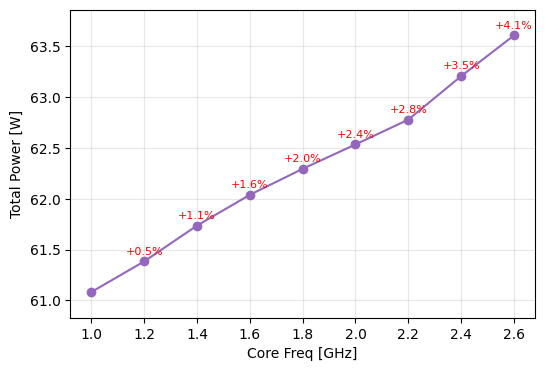}
        \caption{Power vs. Core Freq.}
        \label{fig:memthrash_pow_core}
    \end{subfigure}\hfill
    \begin{subfigure}[t]{0.24\textwidth}
        \centering
        \includegraphics[width=\linewidth]{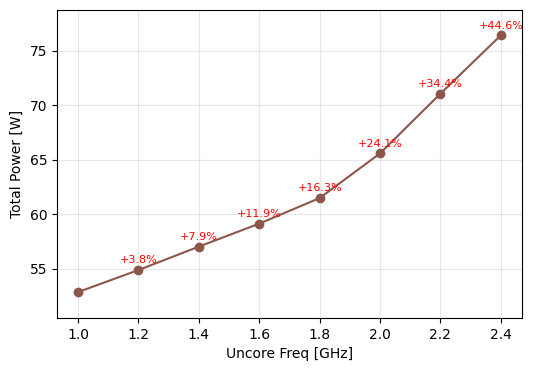}
        \caption{Power vs. Uncore Freq.}
        \label{fig:memthrash_pow_uncore}
    \end{subfigure}

    \caption{Impact of core and uncore frequency scaling on the mixed-workload \texttt{BG\_Memthrash} microservice.}
    \label{fig:Memthrash_core_uncore_impact}
\end{figure*}

Crucially, each dataset incorporates a diverse mix of CPU-bound, memory-bound, and mixed-workload microservices. Each workload class exhibits a unique power and performance profile in response to independent core and uncore frequency scaling. We illustrate the impact of core and uncore frequency scaling on performance (microservice latency) and power consumption (package and DRAM power) across three representative benchmarks: a CPU-bound (\texttt{BG\_CPUAckermann}) in Fig.~\ref{fig:ackermann_core_uncore_impact}, a memory-bound (\texttt{BG\_MemrateFlush}) in Fig.~\ref{fig:MemrateFlush_core_uncore_impact}, and a mixed (\texttt{BG\_Memthrash}) microservice in Fig.~\ref{fig:Memthrash_core_uncore_impact}. To isolate the specific impact of core frequency scaling, we fix the uncore frequency to its minimum. Similarly, we lock the core frequency at its minimum when varying the uncore frequency.

To establish a performance reference for each service, we sweep joint core/uncore frequency configurations and execute the service in isolation on the socket across the defined load levels. We record end-to-end response times and define the target latency for each microservice at each configuration as the 70th percentile of the measured samples.

\begin{figure}[b]
\centering
    \includegraphics[width=0.7\textwidth]{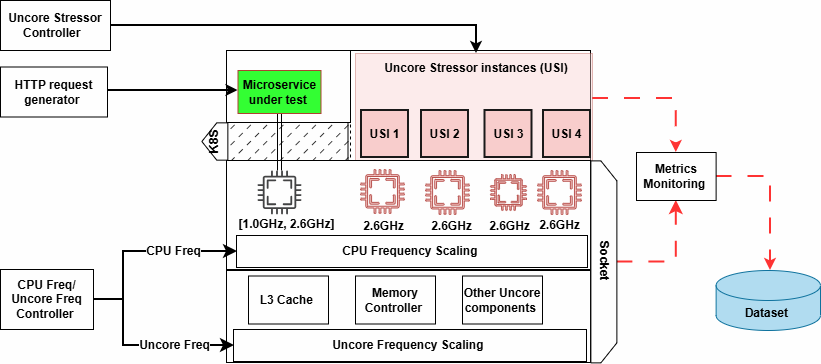}  
    \caption{Uncore Pressure Generation and Its Real-World Impact: Trace Collection Setup}
    \label{fig:Uncore_Interference_Setup}
\end{figure}
\subsubsection{Uncore Interference Generation}
\label{sec:interference_generation}

To evaluate the controller's robustness against noisy neighbors, we generate controlled uncore interference using the PMBW benchmark. We employ two distinct memory-bandwidth stressor profiles: \texttt{ScanRead} (memory read pressure) and \texttt{ScanWrite} (memory write pressure). 

As illustrated in Fig.~\ref{fig:Uncore_Interference_Setup}, on the 14-core test socket, one physical core is strictly dedicated to the target latency-sensitive microservice deployed within the Kubernetes environment. The remaining 13 cores are utilized to host uncore stressor instances (USI) to generate varying levels of interference. These uncore stressors are deployed outside the Kubernetes boundary as standalone, non-containerized processes. To ensure precise and consistent interference scaling, each stressor instance is pinned to a distinct physical core via \texttt{taskset}, allocated 2\,GB of memory, and its core frequency is statically locked at the maximum 2.6\,GHz. By dynamically scaling the number of active stressor instances from 1 to 13, we accurately emulate escalating levels of background uncore contention. This process generates a detailed dataset for every benchmark across our training, evaluation, and test dataset splits. Specifically, for every core and uncore frequency combination, the dataset records both the microservice's performance and the resulting socket-level metrics under varying levels of uncore stress.

\begin{figure*}[t]
    \centering
    
    \begin{subfigure}[t]{0.32\textwidth}
        \centering
        \includegraphics[width=\linewidth]{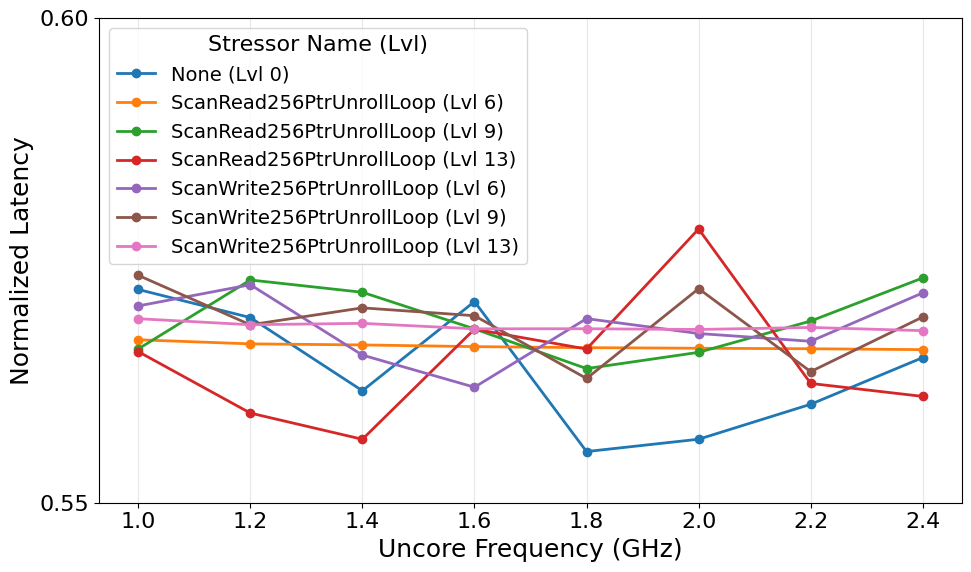}
        \caption{impact on CPU-bound microservice performance.}
        \label{fig:impact_cpu}
    \end{subfigure}\hfill
    \begin{subfigure}[t]{0.32\textwidth}
        \centering
        \includegraphics[width=\linewidth]{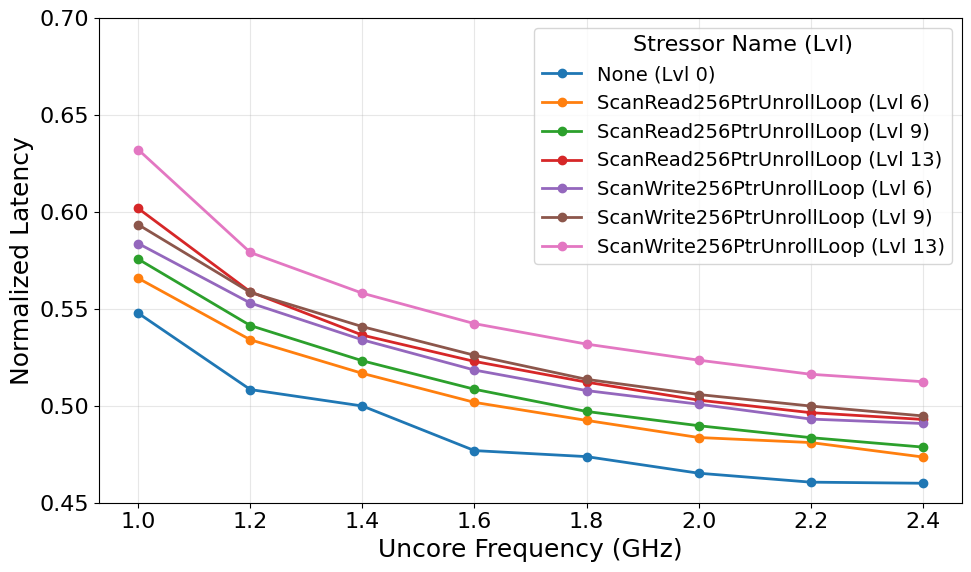}
        \caption{impact on Memory-bound microservice performance..}
        \label{fig:impact_mem}
    \end{subfigure}\hfill
    \begin{subfigure}[t]{0.32\textwidth}
        \centering
        \includegraphics[width=\linewidth]{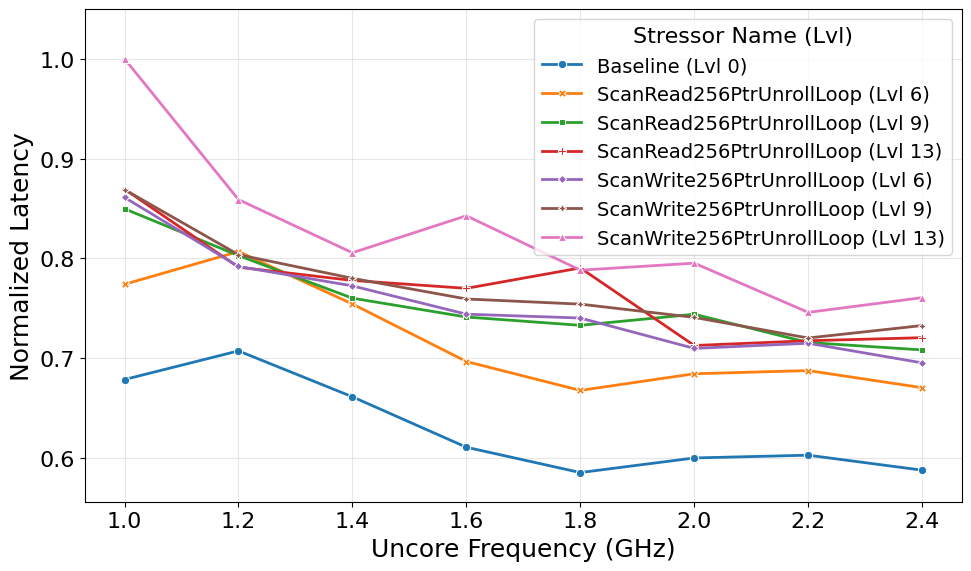}
        \caption{impact on Mixed workload microservice performance.}
        \label{fig:impact_mix}
    \end{subfigure}

    \caption{Impact of escalating uncore stress levels on different microservice workload profiles with core frequency fixed at 1.8\,GHz and maximum load.}
    \label{fig:uncore_impact_row}
\end{figure*}

Fig.~\ref{fig:impact_cpu}, Fig.~\ref{fig:impact_mem}, and Fig.~\ref{fig:impact_mix} show the impact of escalating uncore stress levels (0, 6, 9, and 13 instances) on CPU-bound, memory-bound, and mixed workloads, respectively. For these experiments, we fix the core frequency at 1.8\,GHz and the offered load at its maximum. The data reveals that CPU-bound workloads are largely immune to background uncore contention, as their latency is strictly compute-dominated. Conversely, memory-bound workloads suffer severe, yet approximately linear, performance degradation as the interference scales. Mixed workloads exhibit a moderate, hybrid impact that depends heavily on their active memory bandwidth reliance. Crucially, the latency degradation footprint is dictated not only by the intensity of the interference (i.e., the uncore stress level) but also by the specific workload nature of the uncore stressor (e.g., \texttt{ScanRead} vs. \texttt{ScanWrite}) and the target microservice workload. This demonstrates the complex landscape of real-world cloud environments, where diverse workloads running on the same socket generate highly variable patterns of uncore interference.

Crucially, the RL agent is completely blind to the explicit stressor configuration and active instance count. It must infer the presence and magnitude of external interference indirectly through socket-level telemetry (e.g., shared uncore counters and memory bandwidth saturation). This design forces the agent to dynamically detect and mitigate performance degradation caused by unseen, co-located workloads using strictly observable hardware signals.

\subsection{Offline reward shaping and weight selection.}

\begin{figure}[h]
\centering
\subfloat[\texttt{BG\_CPUAckermann} microbenchmark\label{fig:offline_rewad_cpu}]{
  \includegraphics[width=\linewidth]{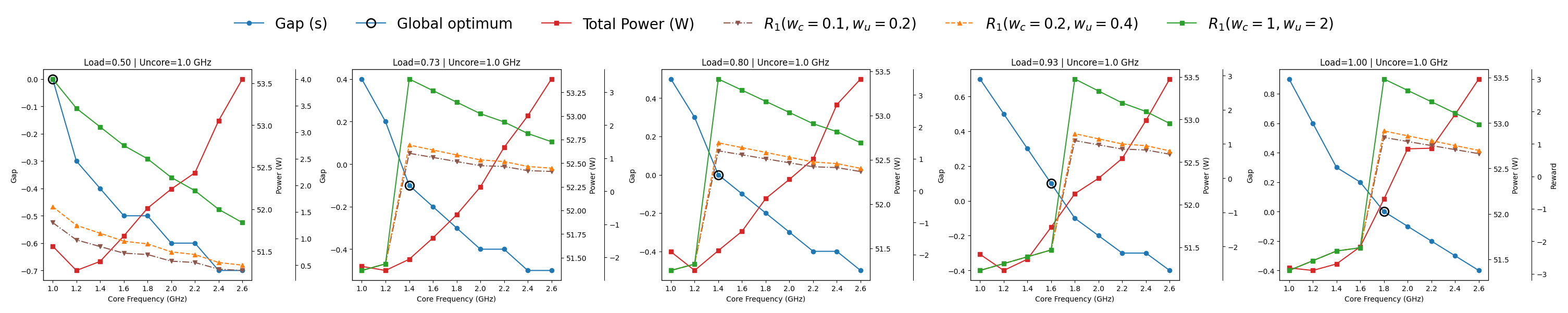}
}\\[-0.2em]
\subfloat[\texttt{BG\_Memcpy} microbenchmark\label{fig:offline_rewad_mem}]{
  \includegraphics[width=\linewidth]{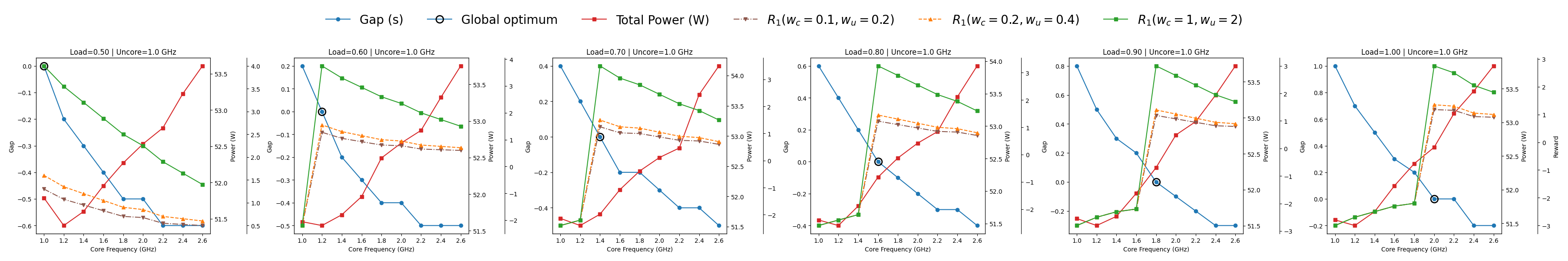}
}
\caption{Reward function behavior under different core and uncore frequency configurations.}
\label{fig:offline_reward_cpu_mem}
\end{figure}

We use historical measurements that jointly record latency outcomes and power consumption over a dense set of core/uncore frequency configurations, spanning different microservices and load levels. For each (microservice, load) condition, we define the \emph{optimal point} as the lowest-power frequency configuration that satisfies the performance requirement (i.e., the minimum-power setting with non-positive gap, $g \le \epsilon$). Reward shaping is calibrated to make this reference point uniquely desirable: the reward attains its maximum at the optimal point, drops sharply under performance requirement violations ($g > \epsilon$), and decreases smoothly (yet perceptibly) in the SLO-safe region when frequencies are unnecessarily high, thereby penalizing wasted power due to over-provisioning.

Because core and uncore frequency scaling have different marginal impacts on package power, we control their relative penalties using weights $(w_c,w_u)$ in $R$. Since the uncore domain is shared and strongly influences package power, we set $w_u = 2 w_c$ (a ratio commonly adopted in vendor and guidance for uncore power management~\cite{juan2012islped}) and evaluate three configurations: $(w_c,w_u)\in\{(0.1,0.2),(0.2,0.4),(1,2)\}$. As shown for \texttt{BG\_CPUAckermann} (see Fig \ref{fig:offline_rewad_cpu}) and \texttt{BG\_Memcpy} (see Fig \ref{fig:offline_rewad_mem}) across multiple loads, $(w_c,w_u)=(1,2)$ yields the clearest shaping: $R$ peaks at the circled global optimum and then decreases approximately linearly as core frequency increases. The smaller-weight settings produce flatter reward curves after reaching the SLO-safe region, which weakens the incentive to downscale. We therefore adopt $(w_c,w_u)=(1,2)$ in the remainder of the study.

\subsection{Impact of State Normalization}
\label{sec:normalization_impact}

We evaluate four normalization strategies:

\begin{itemize}
  \item \textbf{No normalization.} The agent receives raw metrics directly. This introduces zero preprocessing overhead but is prone to severe scale imbalance, where high-magnitude features dominate learning.
  \item \textbf{Global Z-score.} A single Z-score transform is applied to all metrics. Although it standardizes scales, it can distort the physical meaning of key control and target variables (e.g., frequencies and latency).
  \item \textbf{Selective Z-score.} Only background/system metrics are normalized, while control fields are kept in their raw units. This mitigates scale mismatch, but remains frequency-blind by mixing statistics across heterogeneous operating points.
  \item \textbf{Custom (frequency-conditioned).} Separate normalization statistics are maintained for each (core, uncore) frequency pair. This captures regime-specific behavior without cross-contamination, at the cost of a brief cold-start period when a new pair is encountered.
\end{itemize}

\begin{figure}[h]
\centering
    \includegraphics[width=0.7\textwidth]{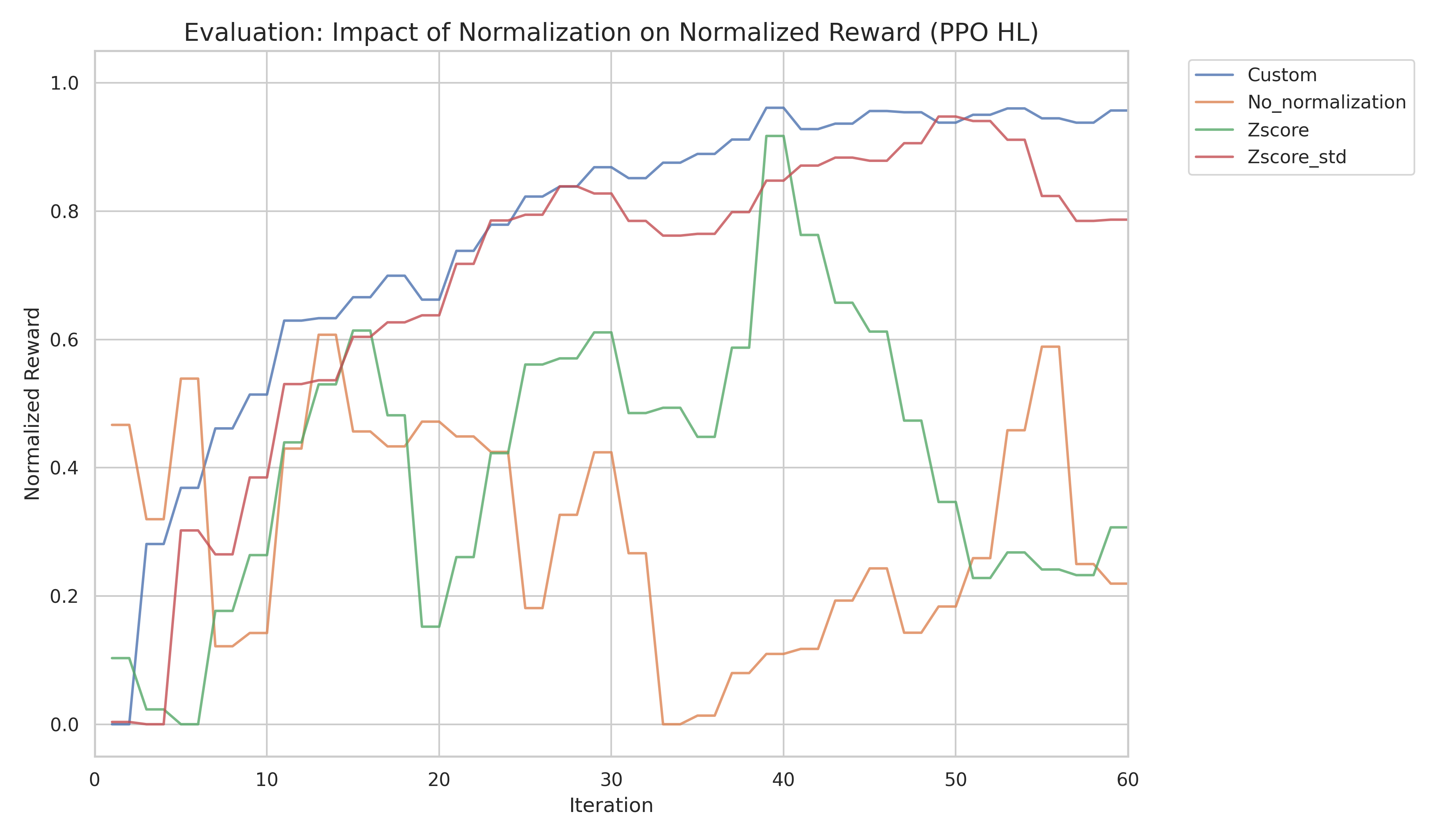}  
    \caption{Normalized reward during evaluation under different normalization strategies.}
    \label{fig:eval_norm_reward}
\end{figure}

The quantitative impact on training stability is summarized in Table~\ref{tab:normalization_impact}.  The performance is defined as the mean normalized evaluation reward after convergence. Figure \ref{fig:eval_norm_reward} shows the normalized reward during evaluation under different normalization strategies. 

\begin{table}[t]
\centering
\caption{Impact of normalization methods on training stability and convergence.}
\label{tab:normalization_impact}
\small
\setlength{\tabcolsep}{5pt}
\renewcommand{\arraystretch}{1.1}
\begin{tabular}{lccc}
\toprule
\textbf{Method} &
\textbf{Convergence (iters)} &
\textbf{Final performance} &
\textbf{Post-conv. CoV} \\
\midrule
\textbf{Custom (freq-conditioned)} & 15--20 & 0.892 & 10.1\% \\
Z-score (selective)               & 25--30 & 0.850 & 14.7\% \\
Z-score (global)                  & 30--40 & 0.820 & 18.9\% \\
No normalization                  & 50+ (unstable) & 0.650 & 43.1\% \\
\bottomrule
\end{tabular}
\end{table}

Overall, the \textit{Custom} normalization yields the best combination of sample efficiency and robustness. By isolating statistical baselines per hardware operating regime, it stabilizes learning and evaluation behavior, converging within 15--20 iterations while achieving the highest final performance (0.892). It also exhibits the lowest post-convergence variability (CoV 10.1\%), compared to the highly unstable no-normalization baseline (CoV 43.1\%).
\subsection{RL Configuration and Training Details}
\label{sec:rl_config}
\begin{figure}[b]
\centering
\subfloat[Reward function trajectories  during training.\label{fig:train_reward}]{
  \includegraphics[width=0.49\linewidth]{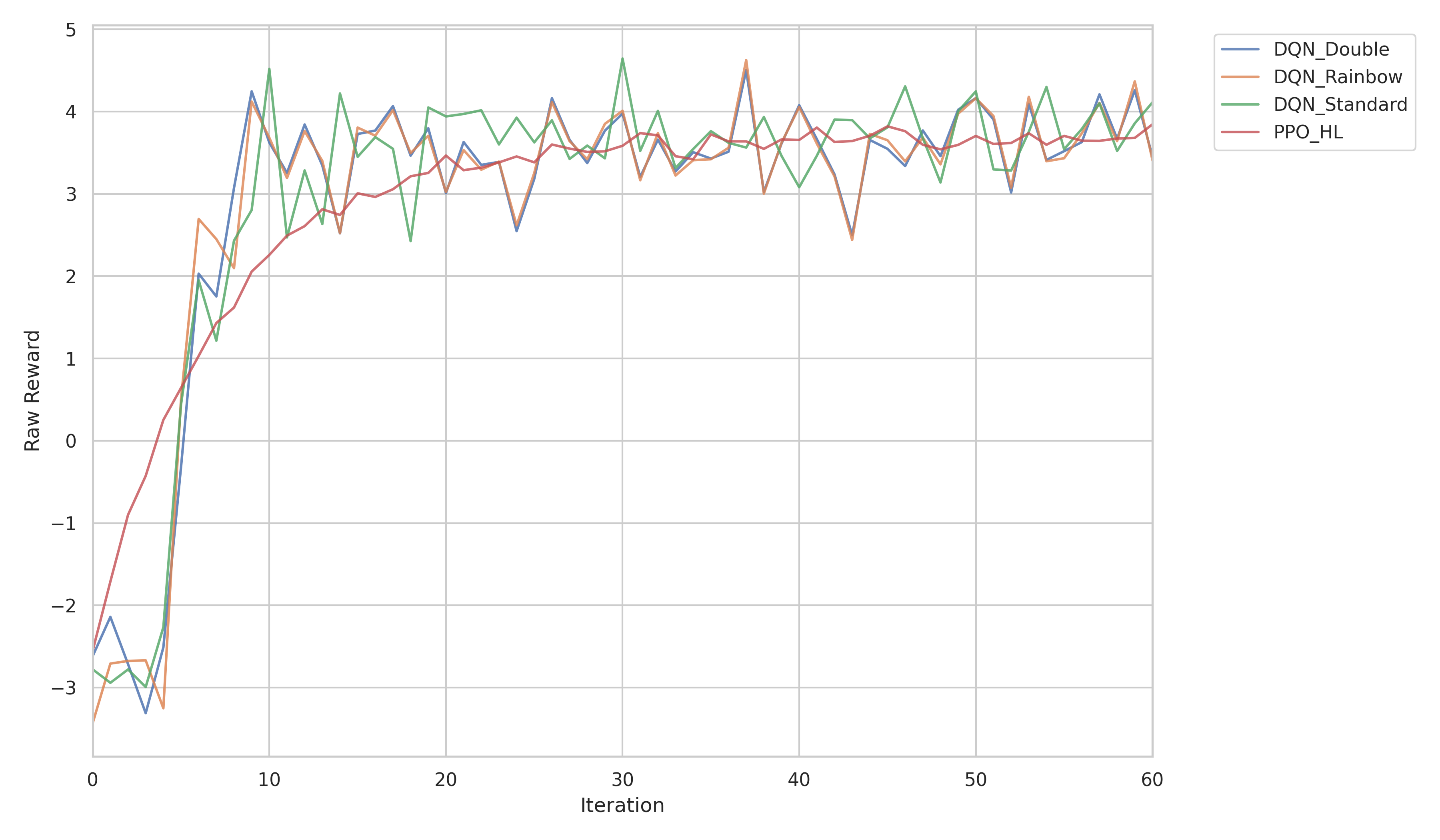}
}
\subfloat[Reward function trajectories  during evaluation \label{fig:eval_reward}]{
  \includegraphics[width=0.49\linewidth]{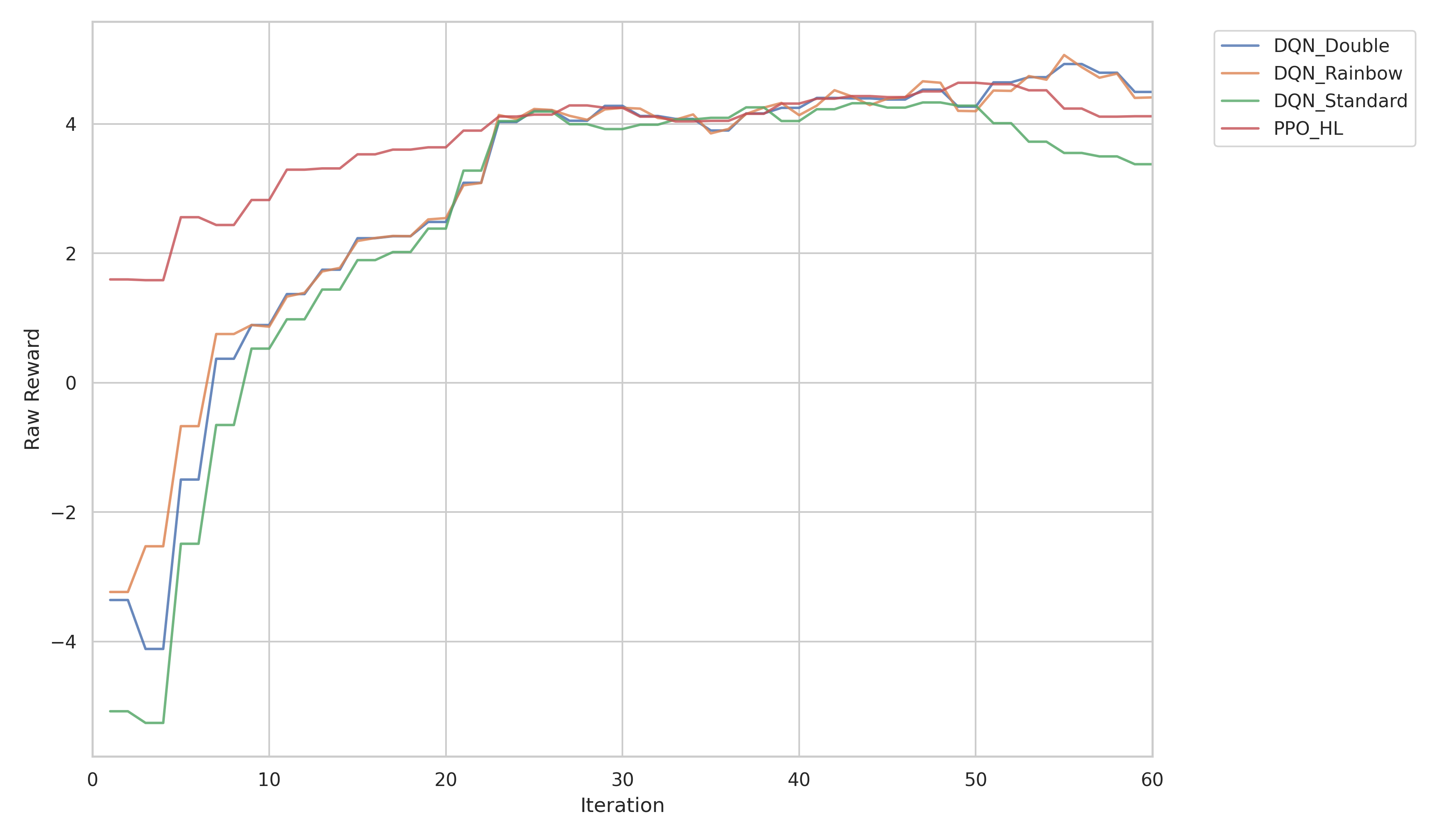}
}
\caption{Reward function trajectories for different RL algorithms}
\label{fig:reward_subfigs}
\end{figure}

\begin{table}[t]
\centering
\caption{Minimal hyperparameter summary. We keep the same training budget ($N_{\text{steps}}$) and shared settings across all methods; only algorithm-specific knobs differ.}
\label{tab:min_hparams_hl}
\footnotesize
\setlength{\tabcolsep}{6pt}
\renewcommand{\arraystretch}{1.12}
\begin{tabular}{lcc}
\toprule
\textbf{Parameter} & \textbf{HL-PPO} & \textbf{HL-DQN} \\
\midrule
Training iteraction & \multicolumn{2}{c}{$N_{\text{steps}}$ =100} \\
Optimizer / $lr$ & \multicolumn{2}{c}{Adam; $3\times 10^{-4}$} \\
Discount $\gamma$ & \multicolumn{2}{c}{0.99} \\
Train batch size & \multicolumn{2}{c}{4096} \\
\midrule
Policy structure & \multicolumn{2}{c}{Hierarchical (coarse+fine)} \\
Action space size & \multicolumn{2}{c}{$(|\mathcal{A}_{\text{coarse}}|,|\mathcal{A}_{\text{fine}}|)=(9,25)$} \\
\midrule
PPO: clip / GAE / entropy & $0.2$ / $0.95$ / $0.01$ & -- \\
DQN: buffer / target update & -- & 65{,}536 / 500 \\
DQN variants (if used) & -- & Double, Dueling, Rainbow (toggles) \\
\bottomrule
\end{tabular}
\end{table}
We implement the hierarchical controller in Ray RLlib and adopt Proximal Policy Optimization (HL-PPO) as the default optimizer; hyperparameters and network settings are summarized in Table~\ref{tab:min_hparams_hl}. We select PPO because its clipped on-policy updates are empirically more stable under the non-stationarity induced by dynamic load and co-runner interference, and it typically requires less sensitive hyperparameter tuning than value-based methods in continuous-control settings.

We validate this choice by comparing HL-PPO with standard DQN, Double DQN, and Rainbow DQN under an identical reward definition and training. Figure~\ref{fig:train_reward} and Figure~\ref{fig:eval_reward} report the reward trajectories during training and evaluation, respectively, across the different RL algorithms. After convergence (approximately iteration $>20$), all methods reach comparable average reward; however, HL-PPO exhibits substantially lower post-convergence reward volatility (about $\pm 0.2$) than Double DQN and DQN (about $\pm 0.5$, $\approx 2.5\times$ higher) and Rainbow DQN (about $\pm 0.6$, $\approx 3\times$ higher). This stability is important for deployment because it translates into more predictable frequency decisions and reduced control jitter.

A further reason for adopting PPO is its gradual on-policy adaptation under workload non-stationarity: in cloud deployments, K8SPI can optionally perform periodic online calibration using newly collected runtime samples to refine the policy. PPO’s clipped updates help limit abrupt policy changes during such fine-tuning, reducing the risk of transient control spikes under distribution shift.

\subsection{Baseline Governors}
\label{sec:baselines}
We benchmark K8SPI against the following baselines:

\begin{enumerate}
    \item \textbf{ \texttt{performance} Governor:} Default used power Governor that pinning core and uncore frequencies to maximum hardware limits, establishing upper bounds for both power and performance.
    
    \item \textbf{Flat Single-Agent PPO (Ablation):} A non-hierarchical PPO agent managing the joint core-uncore action space. For fair algorithmic comparison, it uses K8SPI's exact observation space and a mathematically equivalent single-step reward:
    \begin{equation}
    R_{\text{flat}} = 
    \begin{cases}
    -(2 + g) - \psi, & g > \epsilon \\[0.5em]
    (1 - |g|) + w_c(1 - a_{\text{core}}) + w_u(1 - a_{\text{uncore}}), & g \leq \epsilon
    \end{cases}
    \end{equation}
    
    \item \textbf{Offline Optimal (Oracle):} Computed offline by scanning all frequency permutations per (microservice, load) pair. It selects the tuple $(f_{\text{core}}^{\text{opt}}, f_{\text{uncore}}^{\text{opt}}, P^{\text{opt}}, g^{\text{opt}})$ that strictly satisfies the performance requirement ($g \leq 0$) while minimizing total power, providing an absolute reference to quantify our policy's optimality gap.
\end{enumerate}

\subsection{Performance Metrics}
\label{sec:performance_metrics}
During the training/evaluation phase, the agent is evaluated over entire episodes of length $T$. We define four primary metrics averaged over the episode steps $t \in \{1, \dots, T\}$:

\begin{enumerate}
    \item \textbf{Gap Percentage ($\tilde{g}$):} The mean relative deviation of the resulting latency from the target latency over the episode. Let $g_t$ be the latency gap at step $t$. The mean gap is calculated as:
    \begin{equation}
    \tilde{g} = \frac{1}{T} \sum_{t=1}^{T} g_t
    \end{equation}
    A positive value ($\tilde{g} > 0$) strictly indicates an performance requirement violation, whereas $\tilde{g} \leq 0$ signifies compliance.

    \item \textbf{95th Percentile Gap ($\tilde{g}_{p95}$):} Because the mean gap can obscure transient performance spikes, we also evaluate the 95th percentile gap. This represents the value that 95\% of the observed latency gaps ($g_t$) fall below. It serves as a strict measure of tail latency, ensuring the system remains reliable even under worst-case interference.

    \item \textbf{Power Distance Percentage ($D_P$):} The mean relative deviation of the controller's power consumption ($P_t$) from the offline optimal oracle ($P_t^{\text{opt}}$) across the episode. It is calculated as:
    \begin{equation}
    D_P = \frac{1}{T} \sum_{t=1}^{T} \left( \frac{P_t - P_t^{\text{opt}}}{P_t^{\text{opt}}} \right) \times 100
    \end{equation}
    A lower value means better power efficiency.
    
    \item \textbf{Normalized Reward:} The episodic reward scaled within each testing run to enable fair comparison across different algorithms and configurations.
    
    \item \textbf{Strict Efficiency Score (SES):} A single metric that combines power efficiency and performance requirement compliance, with strong penalties for latency violations. It is defined as:
    \begin{equation}
    \text{SES} = -D_P - \text{Penalty}(\tilde{g})
    \end{equation}
    where $\text{Penalty}(\tilde{g})$ applies a regime-specific cost based on the gap:
    \begin{equation}
    \text{Penalty}(\tilde{g}) = 
    \begin{cases} 
    0, & \tilde{g} \leq -\epsilon \\ 
    \frac{\tilde{g}+\epsilon}{\epsilon}, & -\epsilon < \tilde{g} \leq 0 \\ 
    e^{\frac{\tilde{g}}{\epsilon}}, & \tilde{g} > 0 
    \end{cases}
    \end{equation}
\end{enumerate}

\subsection{Testing Environment}
\begin{figure}[h]
\centering
    \includegraphics[width=0.7\textwidth]{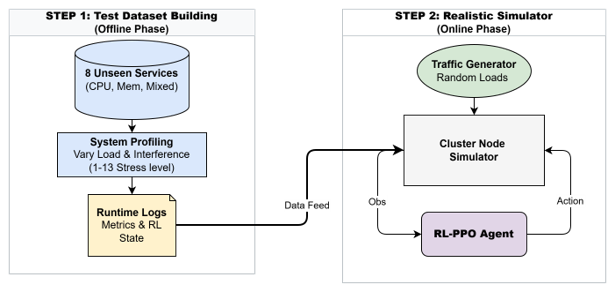}  
    \caption{Normalized reward during evaluation under different normalization strategies.}
    \label{fig:testphase}
\end{figure}
We evaluate the trained policy using a replay-driven simulator (see Figure\ref{fig:testphase} illustrates the testing environment simulator) in closed loop. after the offline phase that builds the test dataset as described in \ref{sec:benchmarks}. The simulator is driven by (i) a traffic generator that determines the time-varying load profile and (ii) a simulator component that replays the system state (online phase).  It emulates a non-stationary cloud node by executing microservices in fixed time windows while varying offered load and uncore interference at a fixed interval, we refer to this fixed interval as a \emph{workload epoch} (\(T_{\text{epoch}}=60\,\)s) to emulate time-varying demand in real cloud settings. At each workload epoch, the PPO agent receives the same observation representation used in the live system and outputs a joint core/uncore frequency adjustment. The simulator resolves each action by replaying the corresponding measured outcome from the dataset, conditioned on the active microservice, load level, interference setting, and selected core/uncore configuration. This design enables reproducible closed-loop testing while avoiding perturbations that would be introduced by exploratory actions on the physical node.

\section{Evaluation and Results}
\label{sec:evaluation}
To thoroughly validate K8SPI's effectiveness, we evaluate the framework across three complex, gradual, real-world scenarios.

\subsection{Scenario 1: Isolated Latency-Critical Microservice}
\label{sec:eval_s1}
\textbf{Setup.} In this scenario, we evaluate our solution in an interference-free environment. A single latency-critical microservice is deployed on the test socket at a time. For each microservice in the test dataset, we run a 300\,s trace under continuous sequential HTTP requests. The load level is updated every 60\,s; (\(T_{\text{epoch}}=60\,\)s). K8SPI employs event-triggered actuation and intervenes only when the performance gap leaves the safety band, i.e., \(g > +20\%\) or \(g < -20\%\).
To evaluate the efficiency of our RL design in reaching optimal hardware configurations with minimal action, we impose strict action budgets. Per load period, the hierarchical controller is allocated a maximum of 6 steps (equivalent to three full coarse$\rightarrow$fine cycles), while the flat single-agent PPO baseline is strictly limited to 3 actions.

\subsection{ Results and Trace Analysis}
\label{sec:eval_s1_results}

\begin{figure}[h]
\centering
    \includegraphics[width=0.7\textwidth]{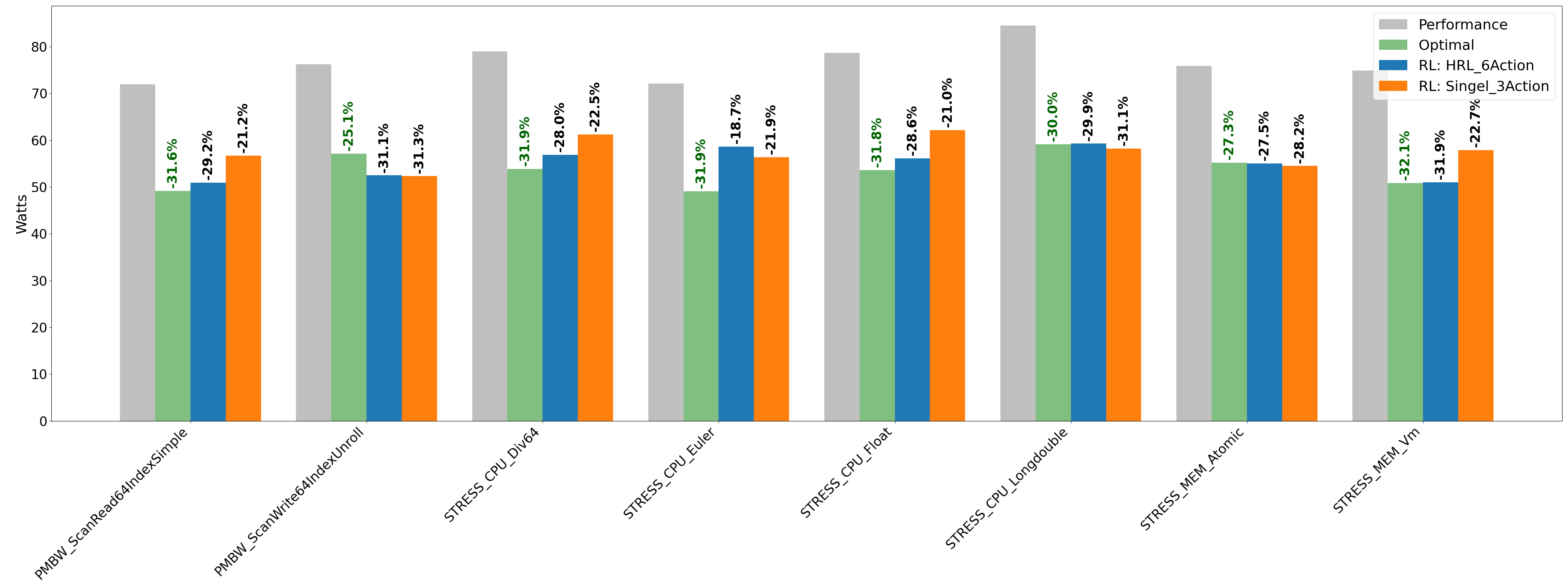}  
    \caption{Total Power Consumption per Benchmark Under Different Power Governance Policies (Scenario 1)}
    \label{fig:power_reduction_summary}
\end{figure}

\begin{figure}[h]
\centering
    \includegraphics[width=0.7\textwidth]{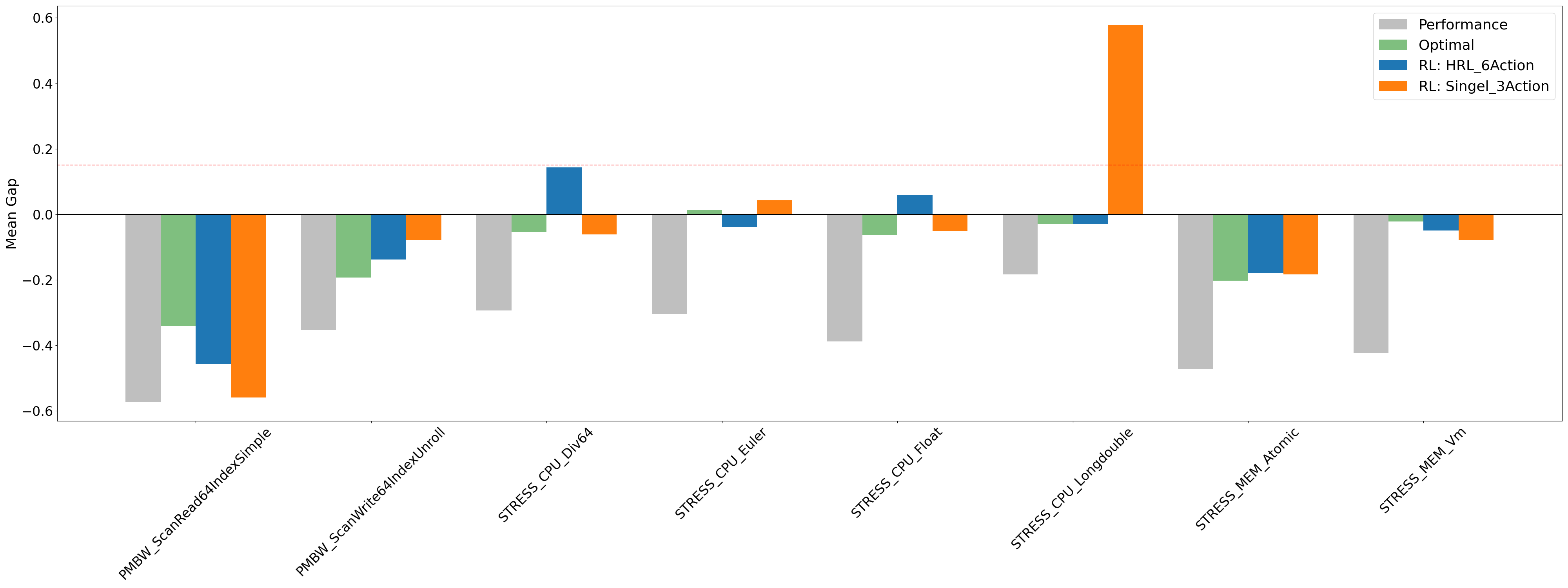}  
    \caption{Average Performance Gap per Benchmark Under Different Power Governance Policies (Scenario 1).}
    \label{fig:performance_gap_summary}
\end{figure}

\begin{figure}[h]
\centering
    \includegraphics[width=0.7\textwidth]{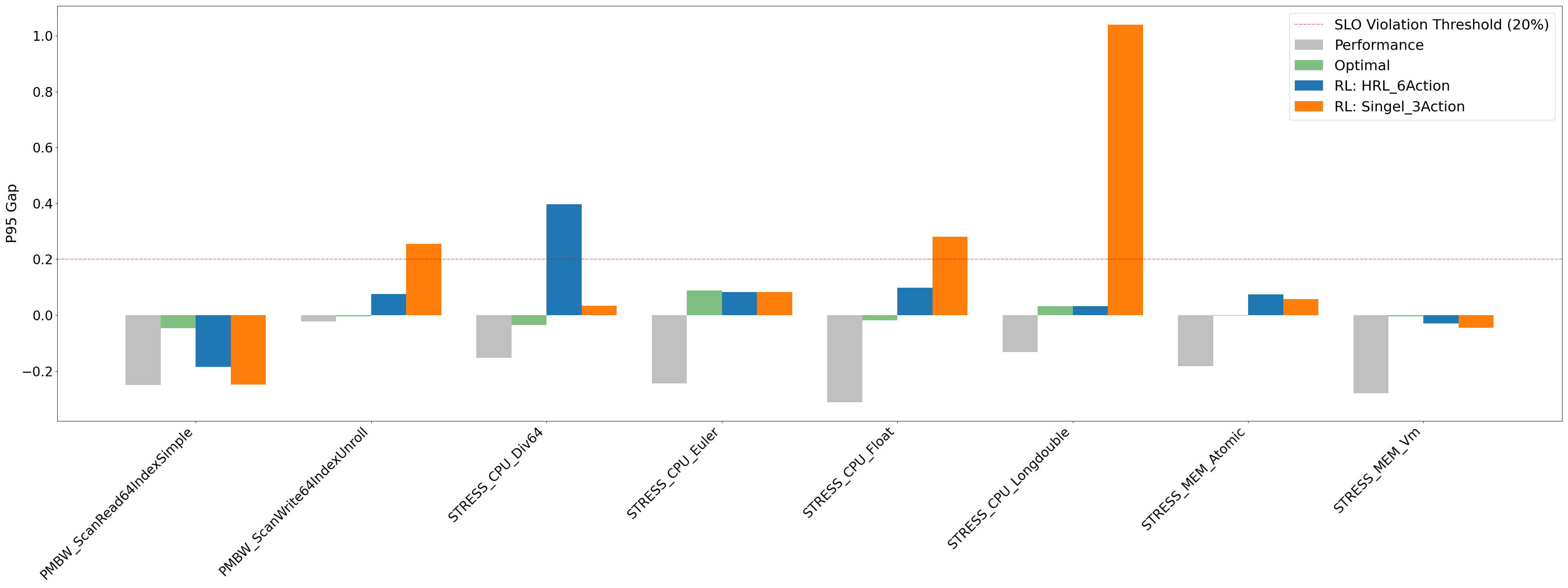}  
    \caption{95th percentile Performance Gap per Benchmark Under Different Power Governance Policies (Scenario 1).}
    \label{fig:performance_p95_summary}
\end{figure}

\begin{figure}[t]
\centering
\begin{subfigure}{0.48\textwidth}
\centering
\includegraphics[width=\linewidth]{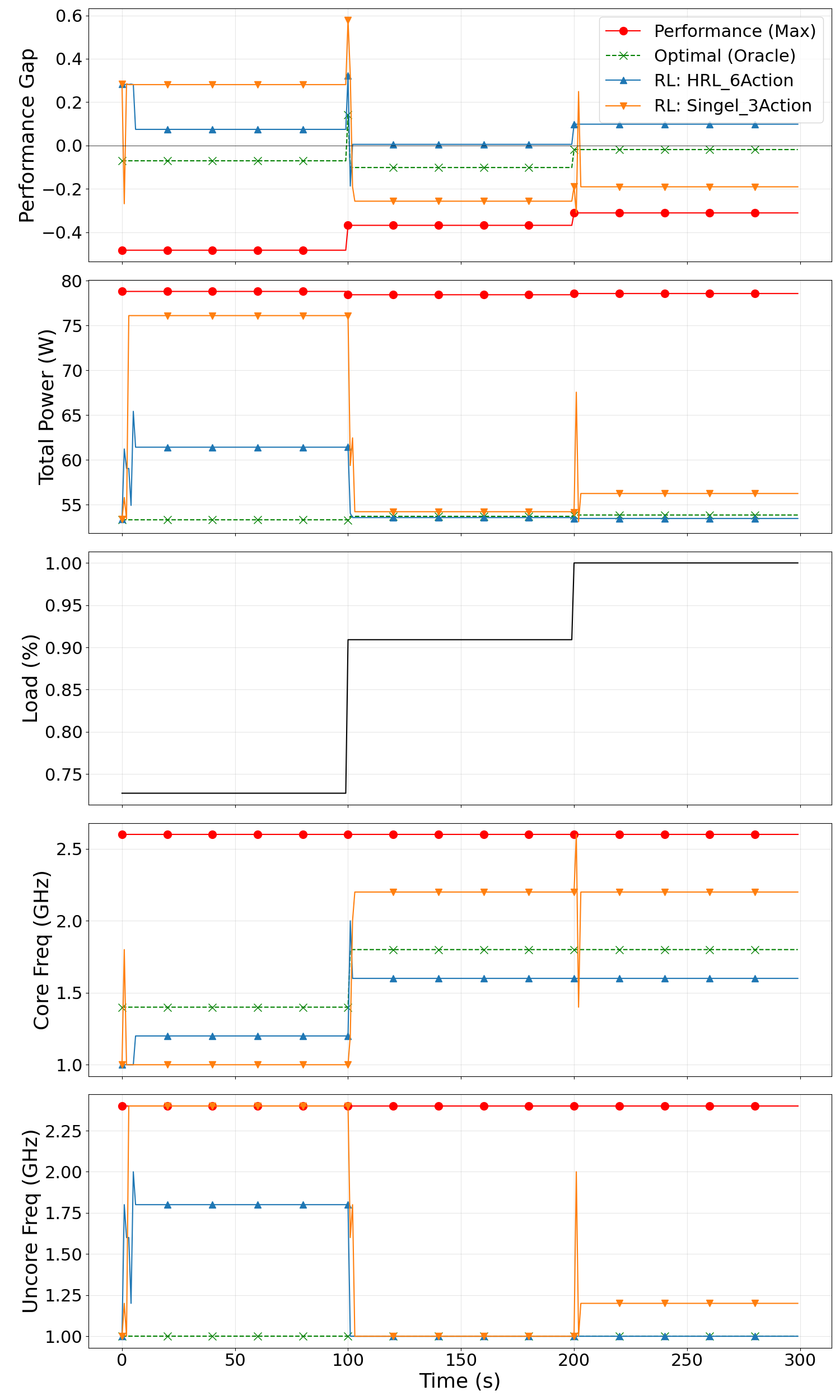}
\caption{\texttt{STRESS\_CPU\_Float}}
\label{fig:STRESS_CPU_Float_combined}
\end{subfigure}
\hfill
\begin{subfigure}{0.48\textwidth}
\centering
\includegraphics[width=\linewidth]{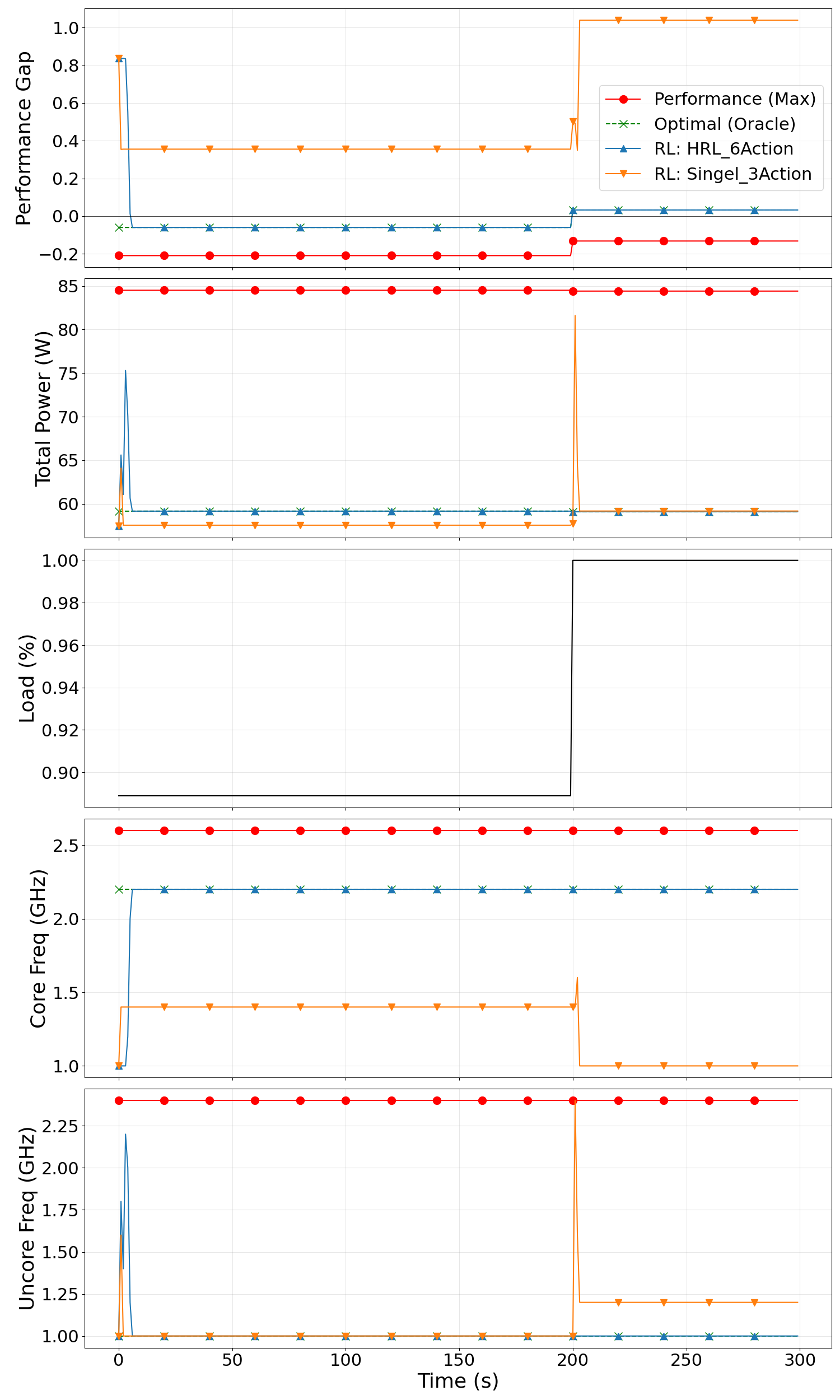}
\caption{\texttt{STRESS\_CPU\_Longdouble}}
\label{fig:STRESS_CPU_Longdouble_combined}
\end{subfigure}
\caption{Trace analysis.}
\label{fig:stress_cpu_trace}
\end{figure}

Across isolated microbenchmarks, the hierarchical controller (HRL) consistently selects safer and more power-efficient operating points than the flat single-agent PPO: as summarized in Fig.~\ref{fig:power_reduction_summary} and Fig.~\ref{fig:performance_gap_summary}, HRL maintains stable performace behavior (gap $\tilde{g}$ near zero or negative) while delivering high power efficiency (typically $\approx$30\% savings) and closely tracking the offline oracle, whereas the single agent achieves only $\approx$20\% savings on average and exhibits instability on sensitive workloads. The clearest failure case is \texttt{STRESS\_CPU\_Longdouble}, where the single agent aggressively downscales and violates the performance requirement ($\tilde{g}=+0.58$ with 100\% violation rate), while HRL preserves compliance ($\tilde{g}=-0.03$) and still attains $\approx$30\% power savings. 

This stability is further validated by analyzing the 95th percentile gap ($\tilde{g}_{p95}$), which exposes transient tail latency spikes that the mean gap might obscure. As depicted in the Fig~\ref{fig:performance_p95_summary}, the flat single-agent PPO frequently breaches the 20\% performance requirement violation threshold ($\tilde{g}_{p95} > 0.2$). It exhibits severe tail latency spikes on workloads such as \texttt{STRESS\_CPU\_Longdouble} ($\tilde{g}_{p95} > 1.0$), \texttt{STRESS\_CPU\_Float} ($\tilde{g}_{p95} \approx 0.28$), and \texttt{PMBW\_ScanWrite64IndexUnroll} ($\tilde{g}_{p95} \approx 0.25$). Conversely, HRL strictly suppresses these worst-case transient spikes, maintaining $\tilde{g}_{p95}$ well below the violation threshold for nearly all evaluated microbenchmarks. While HRL experiences a singular transient spike on \texttt{STRESS\_CPU\_Div64}, it overwhelmingly outperforms the single agent in bounding worst-case latency variations across the dataset.

While aggregate metrics demonstrate HRL's overall efficiency, a trace-level analysis further highlights the performance of HRL compared to the baselines. By examining temporal decisions across distinct microbenchmarks, we identify how HRL successfully navigates the joint core/uncore frequency state-space to prevent destructive local optima.

For the \texttt{STRESS\_CPU\_Float} microservice (See the Fig\ref{fig:STRESS_CPU_Float_combined}), the offered load undergoes discrete step upgrades at $t=100$\,s and $t=200$\,s. During the initial phase, the flat single agent maximizes the uncore frequency to 2.4\,GHz while starving the core frequency at its minimum of 1.0\,GHz. Because this workload is heavily compute-bound, scaling the uncore provides zero latency benefit. Consequently, the single agent consumes 76\,W while continuously violating the $+0.2$ performance requirement threshold (gap $\approx +0.3$). In contrast, HRL correctly identifies the compute sensitivity, stabilizing at a core frequency of 1.2\,GHz and an uncore of 1.8\,GHz. This allows HRL to safely maintain compliance (gap $\approx +0.08$) while saving 15\,W compared to the single agent. As load increases, HRL dynamically readjusts to perfectly overlap the offline oracle's trajectory, pinning the uncore to its minimum and scaling the core to maintain a near-zero gap at optimal power ($\sim$54\,W) without massive oscillations. Meanwhile, at maximum load (200--300\,s), the single agent fails to adapt its actions to the workload's nature, unnecessarily increasing the uncore frequency from 1.0\,GHz to 1.2\,GHz, which needlessly increases total power consumption.

The trace of \texttt{STRESS\_CPU\_Longdouble} (see the Fig\ref{fig:STRESS_CPU_Longdouble_combined}) further exposes the outperformance of the HRL agent compared to the baselines. HRL perfectly mimics the oracle policy throughout the execution (core 2.2\,GHz, uncore 1.0\,GHz), ensuring strict performance requirement compliance (gap never exceeding $+0.02$) at 59\,W. The single agent, however, fundamentally fails to discover an adequate hardware configuration. It operates with a sustained $+0.35$ gap violation initially (0 to 200\,s) and then suffers a catastrophic performance collapse at peak load (gap $> 1.0$) because it inexplicably bottoms out the core frequency to 1.0\,GHz. 

Ultimately, these traces demonstrate that HRL provides a highly generalizable and robust solution for joint core and uncore frequency scaling.

In memory-intensive regimes where core frequency yields diminishing latency benefit, HRL improves energy proportionality by confidently bottoming out compute resources without sacrificing performance: on \texttt{STRESS\_MEM\_Vm}, HRL reaches 51\,W (31.9\% savings) versus 58\,W for the single agent (22.7\% savings), effectively matching the oracle (50.9\,W), and on \texttt{PMBW\_ScanRead64} HRL achieves 29.2\% savings versus 21.2\% for the single agent; more broadly, HRL remains within 1--2\% of the oracle across nearly all memory-bound cases.

Overall, under interference-free conditions, HRL exhibits oracle-level efficiency while maintaining performance-stable behavior, whereas the flat single-agent PPO is more prone to conservative local optima and occasional under-provisioning.

\subsection{Scenario 2: Single Microservice with Best-Effort Interference}
\label{sec:eval_s2_results}

\begin{figure}[!h]

    \begin{minipage}[c]{0.55\textwidth}
        
        \includegraphics[width=\linewidth]{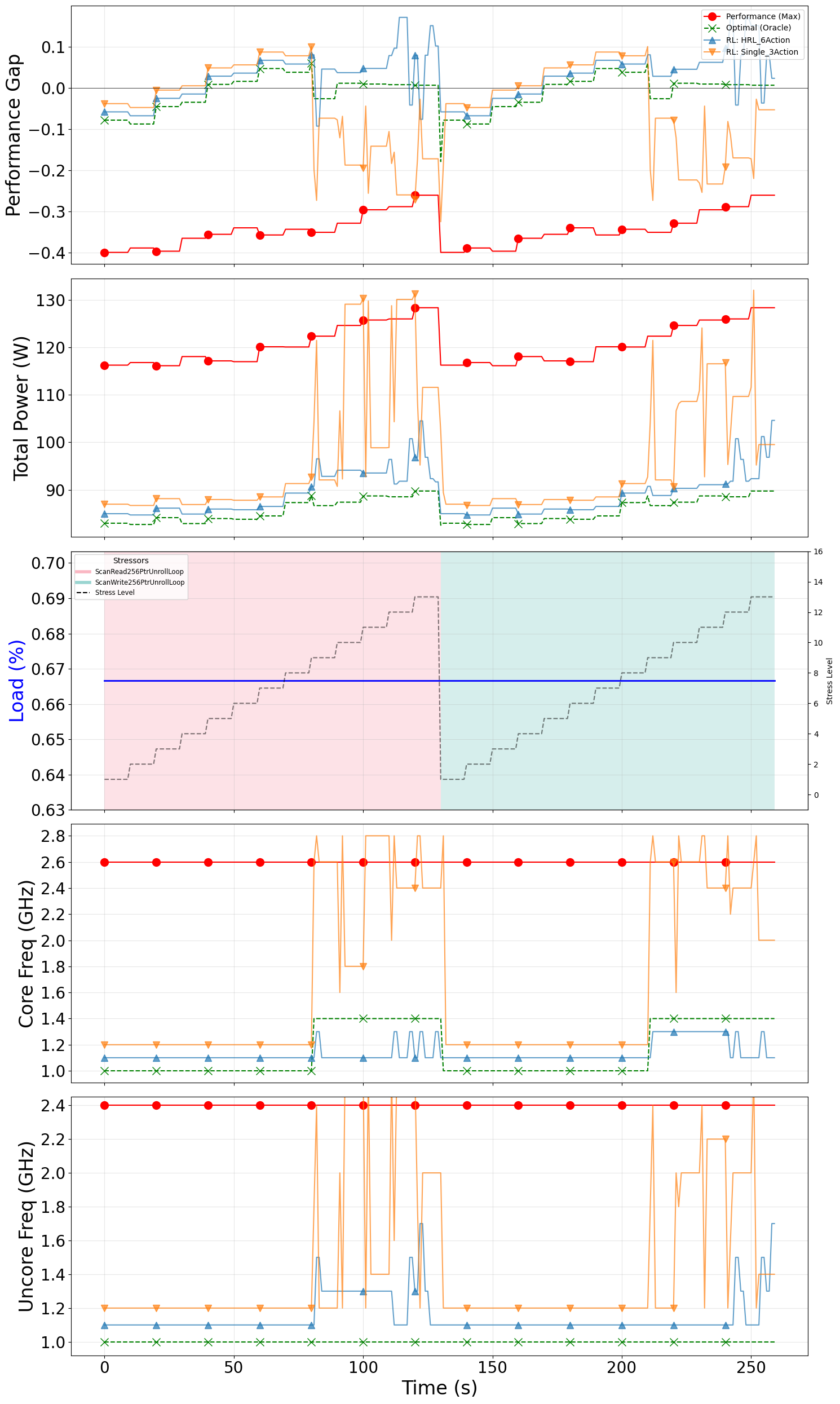}
        \caption{Trace analysis: \texttt{BG\_Memthrash}.}
        \label{fig:s2_memthrash_trace}
    \end{minipage}\hfill
    \begin{minipage}[c]{0.37\textwidth}
        
        \begin{subfigure}{\linewidth}
           
            \includegraphics[width=0.95\linewidth]{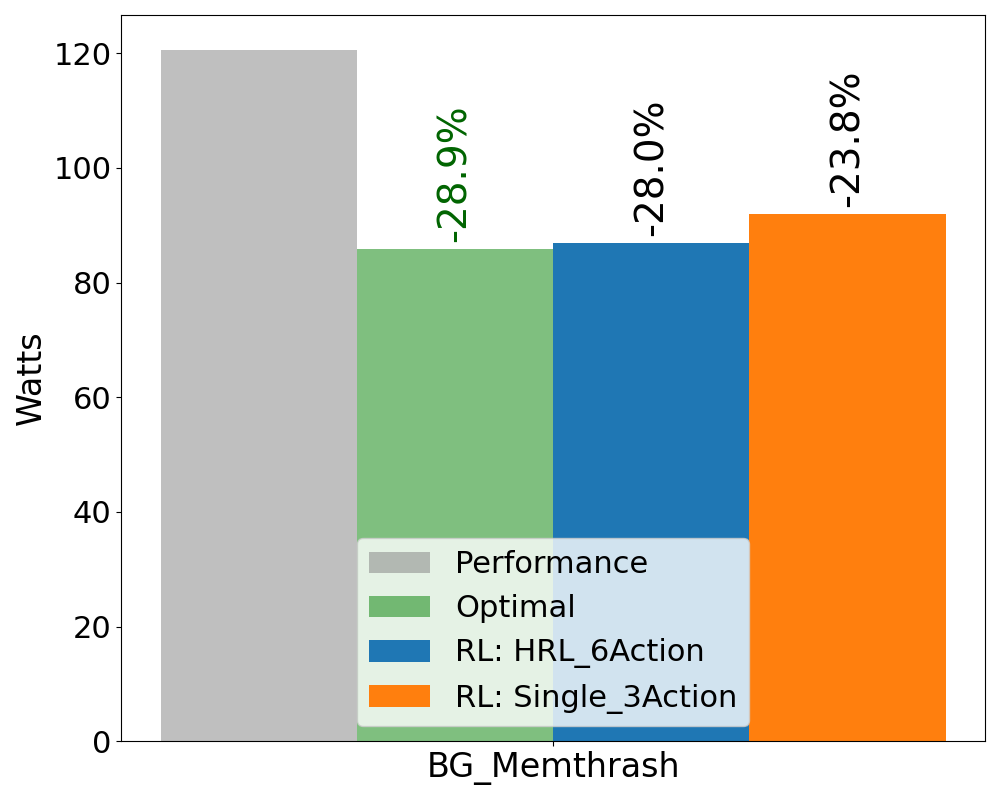}
            \caption{Power reduction summary.}
            \label{fig:s2_power_reduction}
        \end{subfigure}
        
        \vspace{0.5em} 
        
        \begin{subfigure}{\linewidth}
          
            \includegraphics[width=0.95\linewidth]{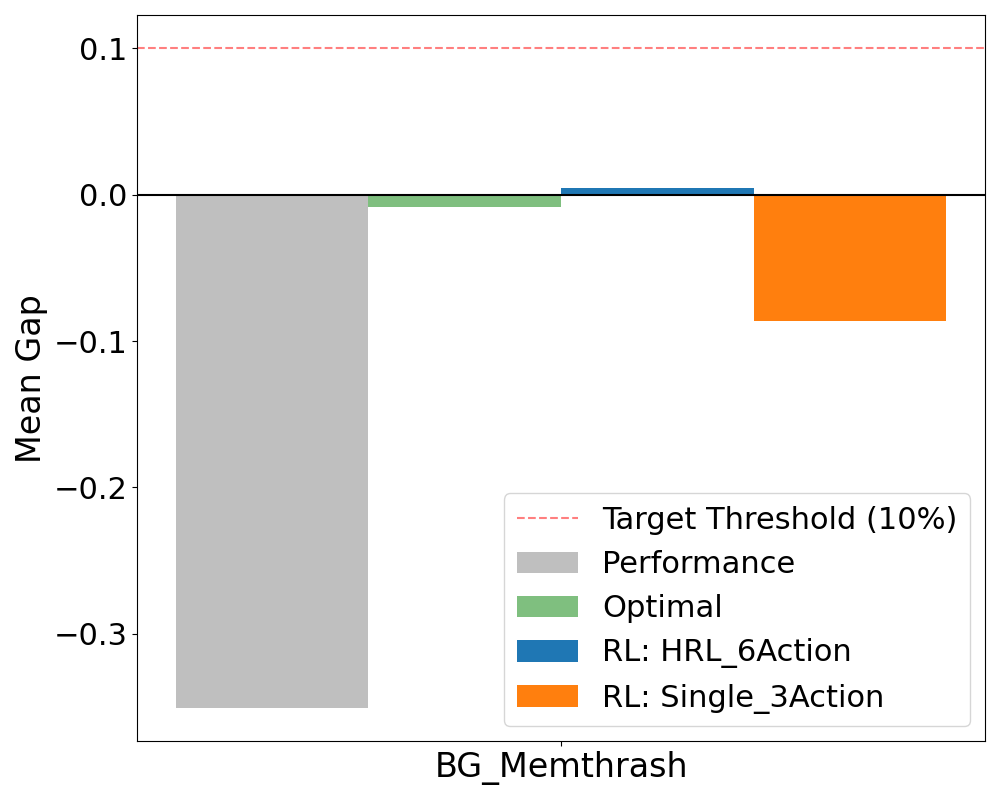}
            \caption{Mean gap summary.}
            \label{fig:s2_gap_summary}
        \end{subfigure}
        
        \vspace{0.5em} 
        
        \begin{subfigure}{\linewidth}
           
            \includegraphics[width=0.95\linewidth]{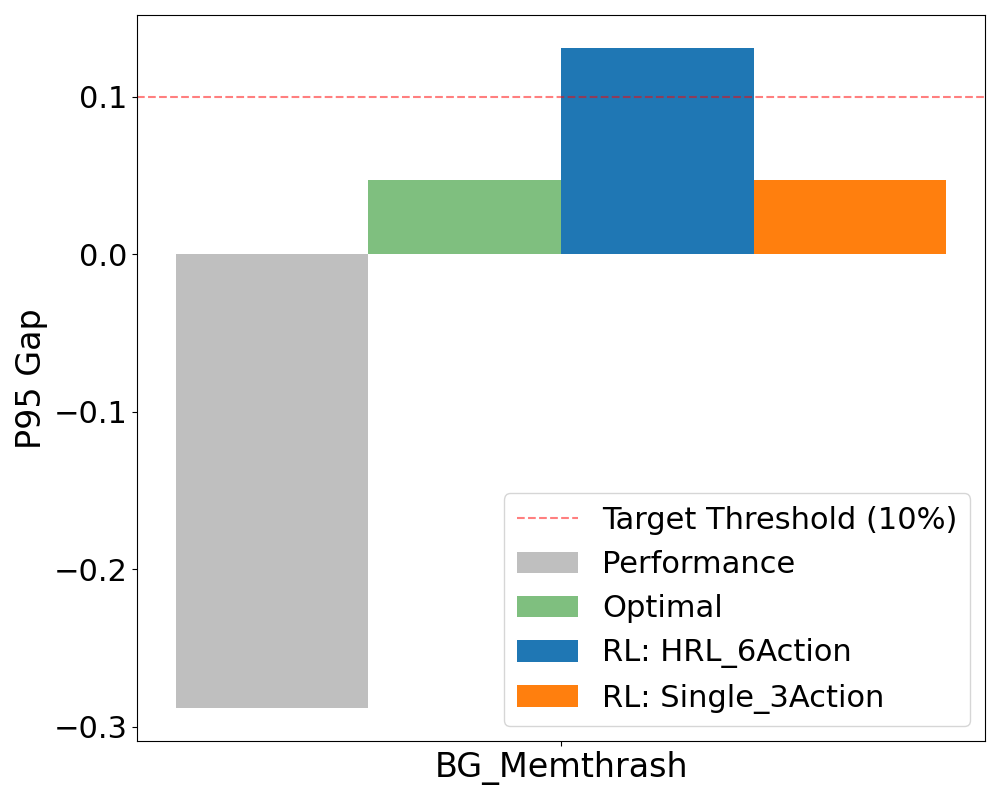}
            \caption{P95 gap summary.}
            \label{fig:s2_p95_summary}
        \end{subfigure}
        
        \caption{Performance summary (power reduction, mean performance gap, and P95 performance gap).}
        \label{fig:s2_aggregate_summaries}
    \end{minipage}
\end{figure}

\subsubsection{Step}
To evaluate the controller under resource contention, we design a memory-bandwidth interference sweep. The target service is a memory-bandwidth-bounded application (\texttt{BG\_Memthrash}) operating at a fixed offered load of 67\%. This service is co-located with memory-intensive best-effort (BE) services (no performance requirement is considered for the BE). 

We progressively increase the number of BE co-runners from 1 to 13 to emulate rising memory bandwidth contention, evaluating two BE workload profiles: \texttt{ScanRead} and \texttt{ScanWrite}. As contention intensifies, the target microservice's performance degrades, which manifests as an increased performance gap. Fig~\ref{fig:s2_memthrash_trace} illustrates the interference sweep setup and the resulting trace behavior as the number of BE co-runners increases.
The RL controller is expected to minimize power consumption while strictly maintaining $\tilde{g}$ below the performance requirement constraint. Driven by observed interference intensity, socket-level metrics, and the real-time performance gap, the controller must dynamically adapt core and uncore frequencies to counteract the BE workload interference and its nature (scan read, scan write).

\subsubsection{ Results and Trace Analysis}
\label{sec:eval_s1_results}
As illustrate on the Figure~\ref{fig:s2_memthrash_trace}, under low to moderate memory contention (Stress Levels 1--8), the performance gap ($\tilde{g}$) remains well within the safe boundary (-0.1 to 0.1). In this regime, both the hierarchical reinforcement learning (HRL) and single-agent controllers exhibit comparable baseline performance, maintaining $\tilde{g}$ between -4\% and +3\% while consuming 84 to 87\,W. However, even in this low-stress environment, HRL demonstrates superior action-space efficiency by reliably pinning the uncore frequency to 1.0\,GHz (note: a slight visual margin is added in the trace plots to prevent the HRL line from obscuring the optimal oracle). This strategy prevents unnecessary DRAM power consumption, whereas the single agent exhibits exploratory and inefficient scaling of uncore frequencies.

As best-effort interference amplifies (Stress Levels 9--13) and threatens the 0.10 (10\%) performance gap threshold. The single agent adopts a reactive, brute-force approach, aggressively boosting core and uncore frequencies to mitigate the rising gap. This leads to severe over-provisioning; for instance, at Stress Level 10, the single agent consumes 108.9\,W---wasting approximately 18\,W compared to HRL---to achieve an excessive negative gap of -21.84\%. Throughout these high-interference phases, the single agent exhibits spiky frequency oscillations and power bursts, indicating unstable control and the creation of an energy-expensive, conservative buffer.

Conversely, HRL demonstrates efficiency-aware control that closely aligns with the optimal oracle, avoiding the panic-driven over-provisioning of the single agent. Instead, it accepts a controlled and acceptable performance degradation (+5.01\% gap at peak Stress Level 13, remaining well below the 10\% threshold) to preserve energy proportionality. align with the optimal policy that accepts a marginal gap increase ($\approx$+0.6\%).

Overall summary metrics for the \texttt{BG\_Memthrash} microservice confirm this efficiency. HRL achieves an average power reduction of 28.0\% relative to the maximum performance baseline, nearly matching the optimal oracle's 28.9\% savings, while the single agent achieves only a 23.8\% reduction as depicted on Fig~\ref{fig:s2_power_reduction}. Furthermore, HRL maintains a mean gap almost perfectly at zero (see Fig~\ref{fig:s2_gap_summary}). While HRL's 95th percentile (P95) gap experiences brief transient spikes slightly above the 0.10 threshold ($\approx$0.13) as shown in Fig~\ref{fig:s2_p95_summary}, the single agent aggressively suppresses these variations at a significant power cost. By reducing overall gap variance ($\sigma_{gap}$) despite stochastic memory stress, the hierarchical architecture successfully navigates the complex power-performance trade-offs inherent to memory subsystem contention.

Overall, for the \texttt{BG\_Memthrash} microservice, HRL achieves a 28.0\% power reduction, nearly matching the oracle's 28.9\% and outperforming the single agent's 23.8\%. HRL maintains an optimal mean gap near zero, whereas the single agent over-provisions to force a negative mean gap of $\approx$-0.09. While HRL's 95th percentile (P95) gap briefly exceeds the 0.10 threshold ($\approx$0.13), the single agent strictly suppresses its P95 gap to $\approx$0.05 at a significant power cost (4.2\% more power). By reducing the gap under stochastic stress, the hierarchical architecture successfully navigates non-linear power-performance trade-offs in the memory subsystem.

\subsection{Scenario 3: Co-located Latency-Sensitive Microservices}
\label{sec:eval_s3}
\subsubsection{Step}
\label{sec:eval_s3}

To evaluate the controller under complex, multi-tenant conditions, we deploy 14 latency-sensitive microservice instances co-located on the same CPU socket,  as detailed in Table~\ref{tab:s3_services}. These instances are instantiated exclusively from the memory-bound and mixed-workload test benchmarks. The experiment is conducted over a fixed duration of 3000\,s, 

To emulate a highly dynamic cloud environment, the offered load for each service changes randomly at the onset of each workload epoch ($T_{\text{epoch}}=60$\,s). Throughout the execution, the co-located services naturally interfere with one another by competing for shared memory bandwidth and uncore resources. Consequently, the performance requirement compliance of each individual instance is simultaneously challenged by its own stochastic load variations and the compounding effects of cross-service interference.

\begin{table}[htbp]
\caption{Scenario 3 Co-located Microservices Configuration}
\label{tab:s3_services}
\centering
\begin{tabular}{|l|c|l|}
\hline
\textbf{Service Type} & \textbf{Instances} & \textbf{Workload Characteristic} \\
\hline
\texttt{PMBW\_ScanWrite256IndexUnroll} & 5 & Memory-bounded (write-intensive) \\
\hline
\texttt{STRESS\_MEM\_Mmap} & 5 & Memory-bounded (mmap operations) \\
\hline
\texttt{PMBW\_PermRead64SimpleLoop} & 4 & Memory-bounded (read-intensive) \\
\hline
\end{tabular}
\end{table}

\subsubsection{Results and Trace Analysis}
\label{sec:eval_s3_results}

\begin{figure}[h]
\centering
    \includegraphics[width=0.7\textwidth]{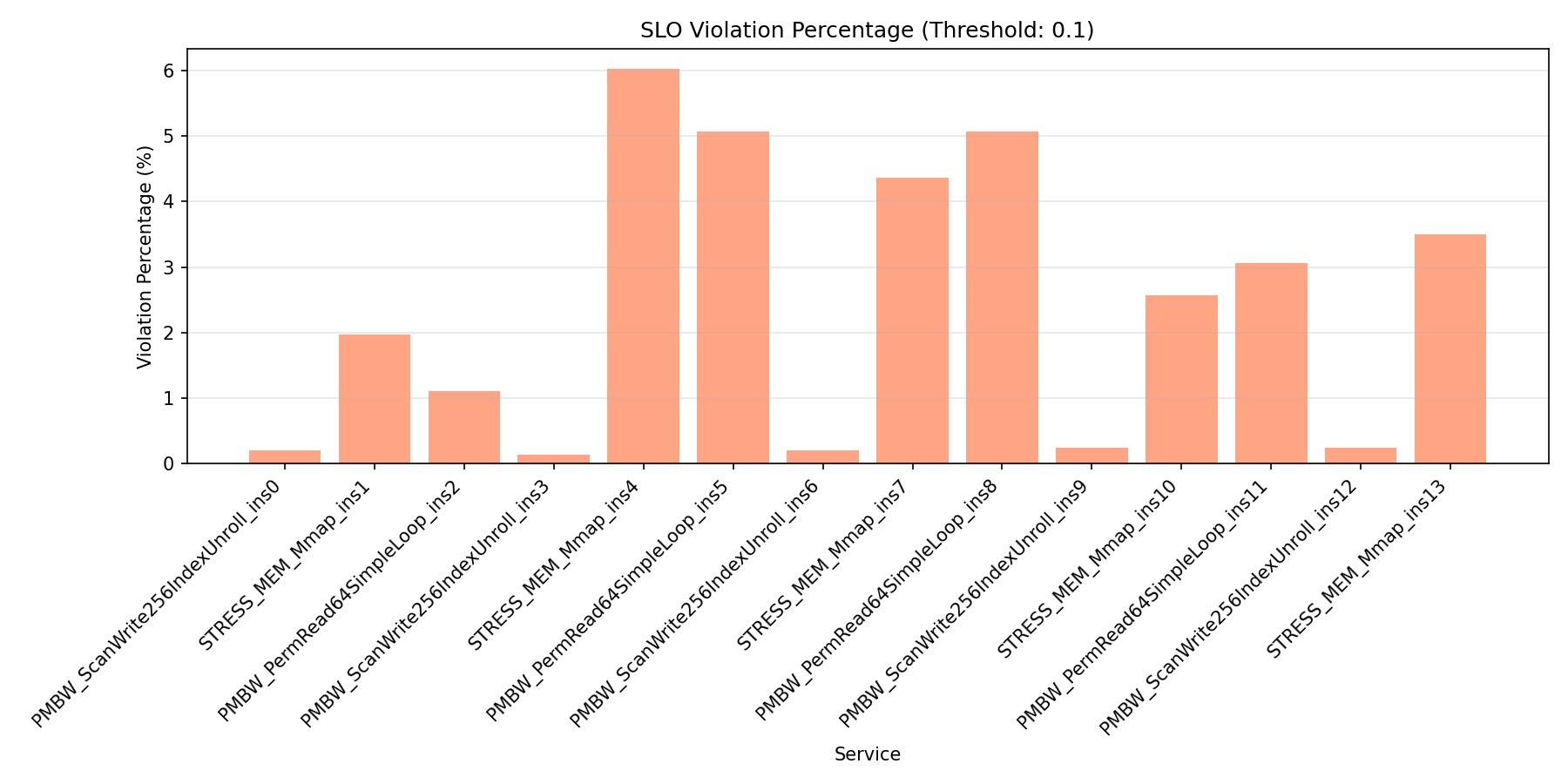}  
    \caption{Detailed breakdown of performance requirement  violation percentages across the diffrent microservices}
    \label{fig:slo_violation_percentage}
\end{figure}

\begin{figure}[h]
    \includegraphics[width=1\textwidth]{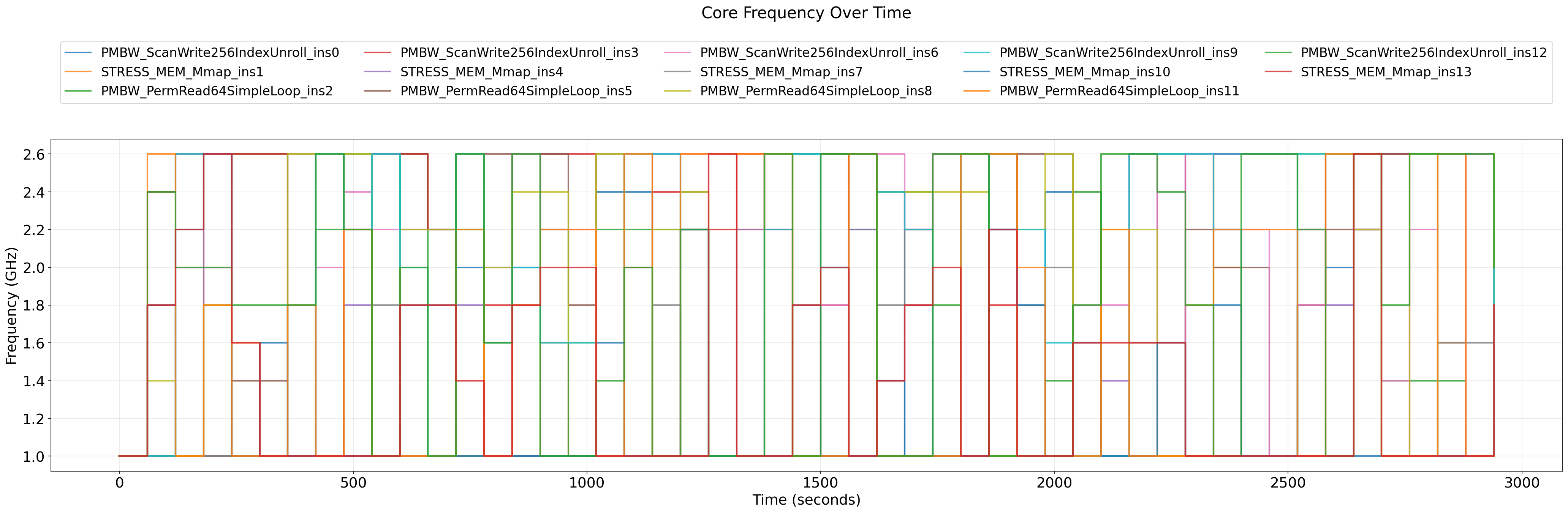}  
    \caption{Dynamic trace of the independent adjustments made to the 14 CPU core frequencies over the 3000-second evaluation.}
    \label{fig:core_freq_vs_time}
\end{figure}

\begin{figure}[h]
\centering
    \includegraphics[width=1\textwidth]{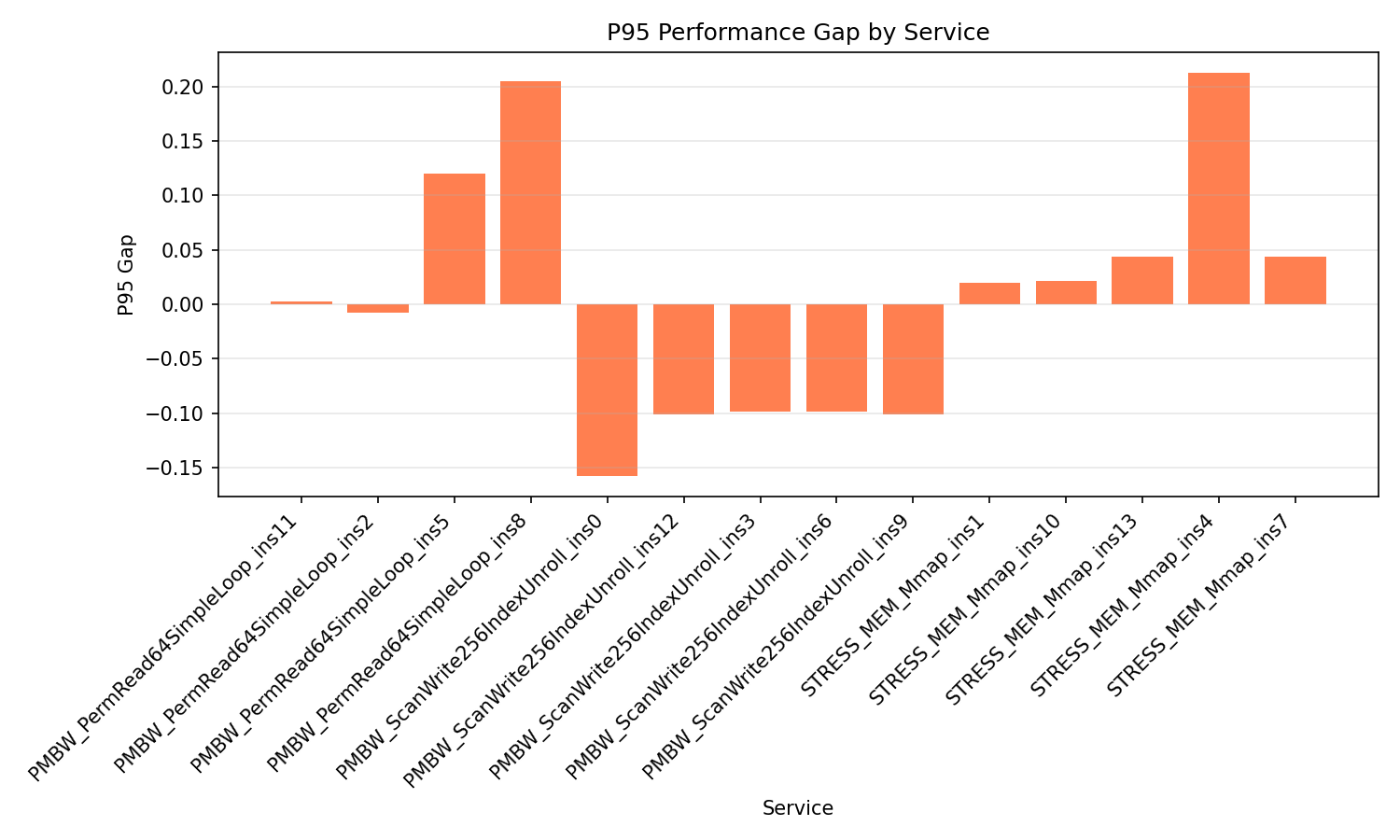}  
    \caption{The $95^{\text{th}}$ percentile (P95) performance gaps evaluated across the deployed microservices.}
    \label{fig:gap_p95}
\end{figure}

\begin{figure}[t]
\centering
\begin{minipage}[t]{0.6\linewidth}
    \centering
    \includegraphics[width=\linewidth]{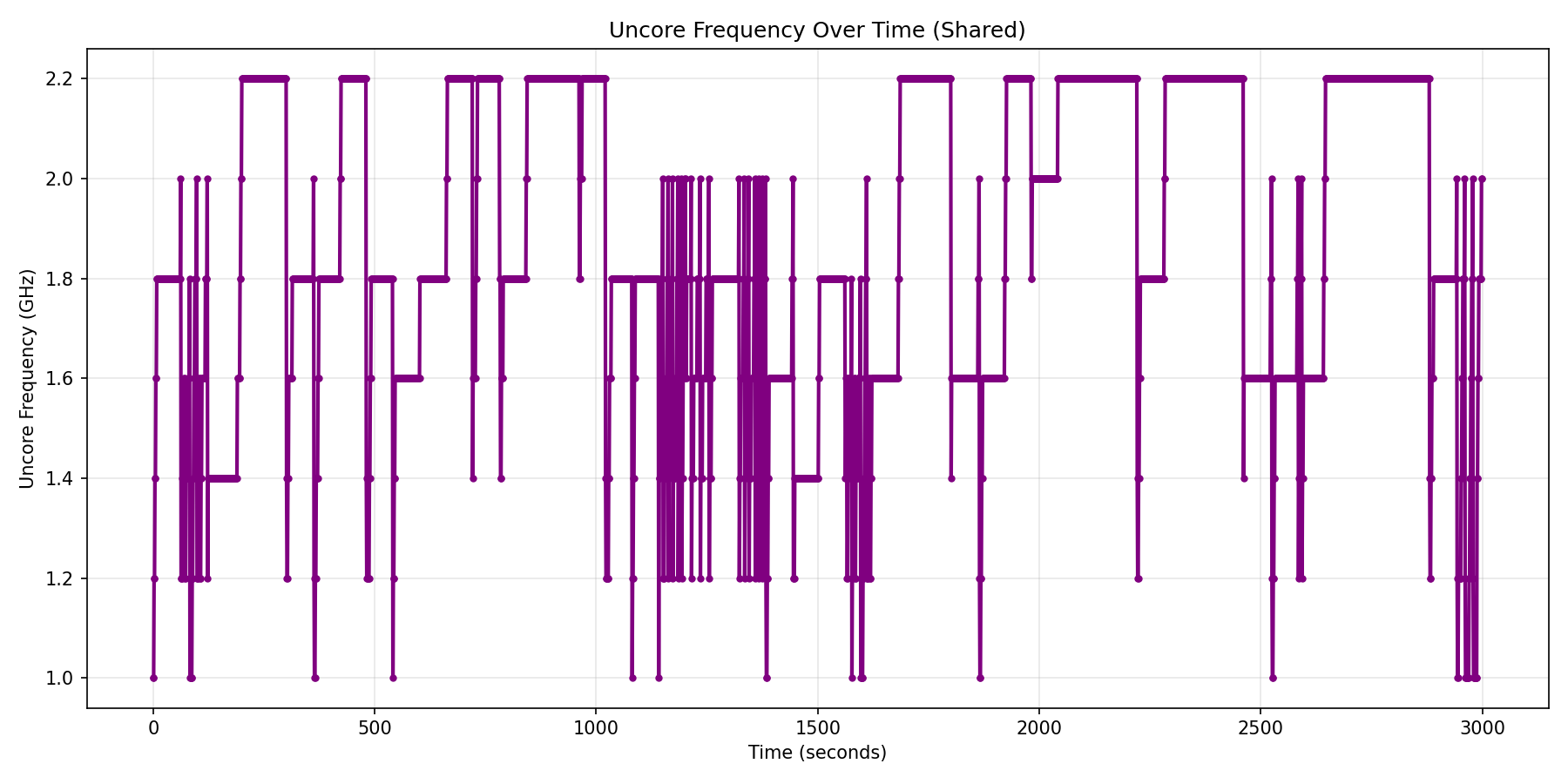}
    \caption{Trace of shared uncore frequency modulations executed by the HRL agent to manage cross-microservice interference.}
    \label{fig:uncore_freq_vs_time}
\end{minipage}
\hfill
\begin{minipage}[t]{0.38\linewidth}
    \centering
    \includegraphics[width=0.7\textwidth]{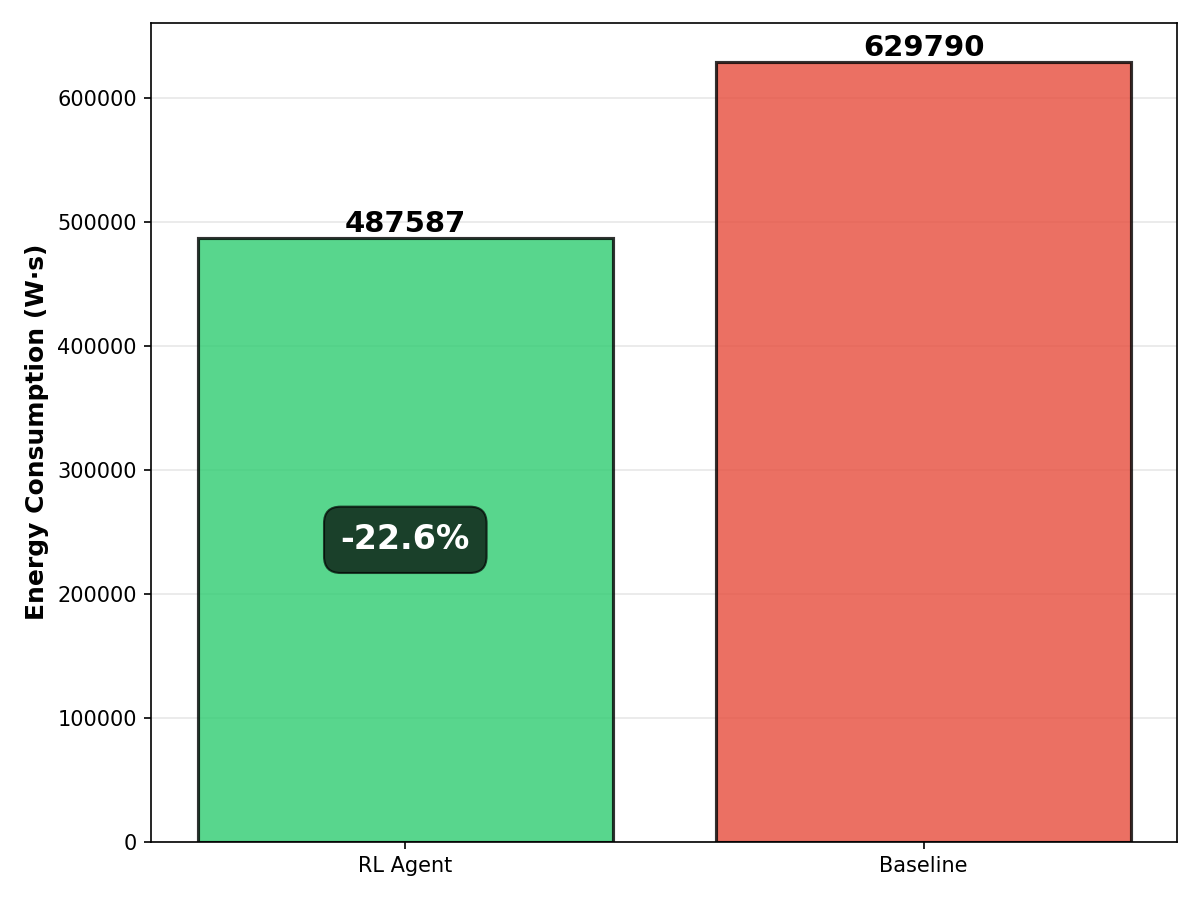}  
    \caption{Total power consumption of  the proposed HRL agent VS the  default performance mode.}
    \label{fig:power_comparison}
\end{minipage}
\end{figure}

Under realistic, multi-tenant cloud conditions where microservices demonstrate stochastic execution behavior, the HRL controller proves exceptionally resilient. The evaluation spans a 3000-second duration. The primary objectives are to enforce a strict performance requirement gap bound of $+0.1$ while maximizing energy savings. Acting entirely autonomously, the HRL agent explores a vast combinatorial space, concurrently managing independent core frequencies across 14 instances alongside the shared uncore frequency. 

Fig. \ref{fig:power_comparison} illustrates the total power consumption reduction achieved by the proposed approach. Overall, the HRL agent achieves a 22.6\% reduction in total power consumption compared to the baseline. Despite stochastic load variations and severe cross-microservice interference at the uncore level, the HRL agent successfully meets the varying performance requirements across the diverse deployed microservices. It achieves an aggregate compliance rate of 97.59\%, suffering only 2.41\% total violations across the entire 3000-second window. 

Fig. \ref{fig:core_freq_vs_time} and Fig. \ref{fig:uncore_freq_vs_time} show the dynamic trace of the independent CPU core and shared uncore frequencies over time. Trace analysis reveals the agent's highly dynamic learned policies: it rapidly and continuously adjusts all 14 independent CPU core frequencies jointly with the shared uncore frequency to maintain strict compliance under highly volatile contention. 

Fig. \ref{fig:slo_violation_percentage} details the violation percentages across the deployed microservices. A detailed breakdown of the violation percentages confirms the policy's robustness. Half of the evaluated microservices---such as \texttt{PMBW\_ScanWrite256IndexUnroll\_ins0}, \texttt{PMBW\_ScanWrite256IndexUnroll\_ins3}, and \texttt{STRESS\_MEM\_Mmap\_ins7}---achieved near-zero ($\approx 1\%$--$2\%$) violation rates over the testing period. Even the most sensitive and demanding microservice, \texttt{STRESS\_MEM\_Mmap\_ins4}, peaked at a mere $\leq 6\%$ violation percentage under maximal contention scenarios. This represents a highly acceptable worst-case operational bound, ensuring no single noisy neighbor leads to the system's full failure.

Fig. \ref{fig:gap_p95} illustrates the P95 performance gaps for the evaluated microservices. Operating exactly on the edge of a performance requirement threshold is risky under stochastic load. The proposed HRL avoids dangerous limit-riding via conservative optimization boundaries. The $95^{\text{th}}$ percentile (P95) gaps remain in negative or near-zero alignment for the vast multitude of services. Even for the most vulnerable tasks, such as \texttt{STRESS\_MEM\_Mmap\_ins4} and \texttt{PMBW\_ScanWrite256IndexUnroll\_ins8}, the $95^{\text{th}}$ percentile spikes stabilize right at the $+0.2$ boundary. This highlights that any latency spikes above the target are rapidly mitigated transients rather than sustained degradations.
\section{Discussion}
\label{sec:discussion}
The evaluation reveals a consistent control pattern: K8SPI's hierarchical reinforcement learning (HRL) separates rapid mitigation of performance requirement violations from fine-grained power minimization, which prevents conservative ``always-max'' operating points under both isolated execution and multi-tenant contention. This section explains the mechanisms behind this behavior and the implications for cloud-native power governance with core/uncore actuation.

\subsection{Decoupling Safety from Optimization}
The failure of the single-agent baseline underscores a critical flaw in flat RL architectures when applied to latency-sensitive environments: the confounding of immediate constraint satisfaction with long-term efficiency. The single agent frequently falls into conservative local optima, aggressively over-provisioning resources at the first sign of performance degradation beyond the requirement. This behavior mimics traditional utilization-driven governors that conflate high resource allocation with system stability. By decoupling these objectives, K8SPI's hierarchical design ensures that the coarse-grained agent strictly handles rapid performance requirement mitigation, freeing the fine-grained agent to safely explore the lower bounds of power consumption. 

\subsection{Navigating the Memory Wall}
A primary insight from the memory-contention scenarios is the physical limitation of independent frequency scaling. While CPU core frequency generally exhibits a linear relationship with computational throughput for CPU-bound tasks, the memory subsystem (governed by the uncore frequency) presents a strict architectural bottleneck. When this "memory wall" is reached under high contention, further increases in core frequency result in diminishing latency benefits while exponentially increasing power consumption. The HRL agent successfully learns this non-linear boundary. It jointly scales core and uncore frequencies to adapt to a saturated memory bus. This indicates that the HRL model implicitly learns the hardware's microarchitectural constraints, avoiding the power-expensive states that trap default governor strategies.

\subsection{Implications for Cloud Power Governance and Limitations}
Two broader implications follow from the evaluation. First, application-level performance feedback (via the performance gap) is critical: without it, node-level governors cannot distinguish safe power reductions from under-provisioning, especially under interference. Second, uncore scaling governors must be integrated with core DVFS into a single policy. Because each workload has a unique footprint, joint core and uncore governance leverages multi-dimensional optimization to efficiently manage shared resources.

Despite these advantages, the current K8SPI framework has notable limitations. The system's responsiveness is bottlenecked by the observation latency of the monitoring stack. In our evaluation, telemetry aggregation via Prometheus introduces a default observation delay of up to 30\,s. Consequently, making immediate reactions to highly dynamic load shifts and uncore interference remains challenging. To accommodate this, we fixed the load periods to 60\,s ($T_{\text{epoch}}=60$\,s) to allow the observation stack to reliably capture the necessary telemetry across all levels (application latency and hardware performance counters). Deploying K8SPI in highly volatile production environments will require integrating a low-latency, real-time observation system.

Furthermore, the current design is tailored for strict CPU pinning scenarios, utilizing a one-to-one mapping between microservice instances and physical CPU cores. While this isolates core-level compute, future work must extend the framework to cover broader, unpinned CPU execution scenarios and multi-core allocations to fully address diverse cloud-native deployments.

\section{Conclusion and Future Work}
\label{sec:conclusion}

This paper presented K8SPI (Kubernetes Power Irrigation), an event-triggered hierarchical reinforcement learning (HRL) controller for node-level power optimization of latency-sensitive microservices in Kubernetes. K8SPI adopts a two-stage architecture that separates rapid mitigation of performance requirement violations from fine-grained power minimization, enabling stable control over complex power--performance trade-offs. This separation is especially important in consolidated multi-tenant deployments, where co-located microservices have heterogeneous workload footprints and distinct performance requirements while contending for shared uncore resources.

To make RL-driven control practical for real systems, we also developed a rapid prototyping and training workflow. By combining RLlib's parallel training with system-state snapshots and trace-driven replay, the workflow supports efficient multi-environment training using real execution traces, reducing iteration time when reward shaping and policy design are refined.

Evaluation on a Kubernetes testbed shows that K8SPI reduces node-level power by 23\%--30\% relative to the Linux performance governor while keeping performance requirement violations below 2\%--3\%, even under dynamic load fluctuations and severe multi-tenant uncore contention. These results indicate that hierarchical RL with joint core/uncore actuation can avoid conservative performance-mode provisioning and achieve substantial power savings without compromising required performance.

Future work will proceed along two directions. First, we will extend K8SPI beyond strict CPU pinning to support unpinned execution and dynamic multi-core allocations, improving generality across cloud-native scheduling regimes. Second, we will address observability latency by integrating a low-latency telemetry path for critical signals, enabling faster reaction to micro-bursts and short-lived interference events.
\bibliographystyle{IEEEtran}
\bibliography{Ref}

\end{document}